\documentclass{aa}
\usepackage{graphicx}
\usepackage{amssymb}
\usepackage{booktabs}
\usepackage{natbib}
\bibpunct{(}{)}{;}{a}{}{,} 
\usepackage{longtable}
\usepackage{lscape}
\usepackage{rotating}
\usepackage{supertabular}
\usepackage{placeins}
\usepackage[usenames,dvipsnames]{xcolor}
\usepackage{hyperref}  
\hypersetup{linkcolor=blue,citecolor=blue,filecolor=black,urlcolor=blue} 

\usepackage[varg]{txfonts}

\newcommand{\figiac}[1]{Fig. \ref{#1}}

\newfont{\gwpfont}{cmssq8 scaled 1000}

\newcommand{\ccor}[1]{\textcolor{black}{#1}}
\newcommand{\ccob}[1]{\textcolor{black}{#1}}

\def\Mv {M_{500}}
\def\Mv {M_{500}}

\def\Rvyx {R_{500}^\mathrm{Y_{X}}}

\def\Mvyx {M_{500}^\mathrm{Y_{X}}}

\def\Mvher500{M^\mathrm{HE}}

\def\planck{{\it Planck}}
\def\chandra{{\it Chandra}}
\def\amas{{PSZ2 G282.28+49.94}}
\usepackage{color}

\newcommand {\ks} {km~s$^{-1} \;$}

\newcommand {\kss} {km~s$^{-1}$}
\newcommand {\h} {Mpc$\;$}
\newcommand {\ml} {$M_{\odot}/L_{\odot} \;$}

\newcommand{\degree}{\ensuremath{\mathrm{^\circ}}}
\newcommand{\arcm}{\ensuremath{\mathrm{^\prime}\;}}
\newcommand{\arcs}{\ensuremath{\arcmm\hskip -0.1em\arcmm \;}}
\newcommand{\arcmm}{\ensuremath{\mathrm{^\prime}}}
\newcommand{\arcss}{\ensuremath{\arcmm\hskip -0.1em\arcmm}}

\newcommand {\kpc} {kpc $\;$}

\begin{document}
\title{PSZ2 G282.28+49.94, a recently discovered analogue of the famous Bullet Cluster}
\author{I. Bartalucci\inst{1}, M. Rossetti\inst{1}, W. Boschin\inst{2,3,4}, M. Girardi\inst{5,6}, M. Nonino\thanks{We dedicate this paper to the memory of our friend and colleague Mario.}\inst{6}, E. Baraldi\inst{7}, M. Balboni\inst{1,8}, D. Coe\inst{9}, S. De Grandi\inst{10}, F. Gastaldello\inst{1}, S. Ghizzardi\inst{1}, S. Giacintucci\inst{11}, C. Grillo\inst{1,7}, D. Harvey\inst{12}, L. Lovisari\inst{1}, S. Molendi\inst{1}, T. Resseguier\inst{13}, G. Riva\inst{1,7},  T. Venturi\inst{14},  A. Zitrin\inst{15}}
\authorrunning{I. Bartalucci}
\institute{
INAF, IASF-Milano, via A. Corti 12, I-20133 Milano, Italy \label{milano_iasf} 
\and Fundaci\'on Galileo Galilei - INAF (Telescopio Nazionale Galileo), Rambla Jos\'e Ana Fern\'andez Perez 7, E-38712 Bre\~na Baja (La Palma), Canary Islands, Spain 
\and Instituto de Astrof\'{\i}sica de Canarias, C/V\'{\i}a L\'actea s/n, E-38205 La Laguna (Tenerife), Canary Islands, Spain 
\and Departamento de Astrof\'{\i}sica, Univ. de La Laguna, Av. del Astrof\'{\i}sico Francisco S\'anchez s/n, E-38205 La Laguna (Tenerife), Canary Islands, Spain 
\and Dipartimento di Fisica dell'Universit\`a degli Studi di Trieste - Sezione di Astronomia, via Tiepolo 11, I-34143 Trieste, Italy 
\and INAF - Osservatorio Astronomico di Trieste, via Tiepolo 11, I-34143 Trieste, Italy 
\and Dipartimento di Fisica, Università degli Studi di Milano, via Celoria 16, I-20133 Milano, Italy 
\and DiSAT, Università degli Studi dell’Insubria, via Valleggio 11, I-22100 Como, Italy 
\and Space Telescope Science Institute, Baltimore, MD, USA 
\and INAF - Osservatorio Astronomico di Brera, via E. Bianchi 46, I-23807 Merate (LC), Italy 
\and U.S. Naval Research Laboratory, 4555 Overlook Avenue SW, Code 7213, Washington, DC 20375, USA 
\and Lorentz Institute, Leiden University, Niels Bohrweg 2, Leiden, NL-2333 CA, The Netherlands 
\and Center for Astrophysical Sciences, Department of Physics and
Astronomy, The Johns Hopkins University, 3400 N Charles St.,
Baltimore, MD 21218, USA. 
\and INAF—Istituto di Radioastronomia, via Gobetti 101, I-40129 Bologna, Italy 
\and Department of Physics, Ben-Gurion University of the Negev, P.O.
Box 653, Be’er-Sheva, 84105, Israel. 
}

\date{Received  / Accepted}

\abstract{We present a detailed study of the gas and galaxy properties of the cluster \amas\ detected in the \planck\ all-sky survey. 
The intracluster medium (ICM) of this object at z=0.56 exhibits a cometary-like shape. Combining \chandra\ and TNG observations, we characterised the spatially resolved thermodynamical properties of the gas and the spatial and velocity distribution of 73 galaxy members. The cluster structure is quite complex with an elongated core region
    containing the two brightest cluster galaxies and one dense
    group to the south-east. Since there is no velocity difference between
    the core and the south-east group, we suggest the presence of a
    merger along the plane of the sky. This structure is related to
    complex X-ray and radio features, and thus the merger has likely been
    caught during the post-merger phase.
    Comparing the distribution of the ICM and of member galaxies, we find a large offset of $\sim 350$ kpc between the position of the X-ray peak and the centre of a concentration of galaxies, preceding it in the likely direction of motion.  This configuration is similar to the famous Bullet Cluster, leading us to dub \amas\ the ‘\planck\ bullet’, and represents an ideal situation to provide  astrophysical constraints to the self-interaction cross-section ($\sigma/m$) of dark matter particles.
These results illustrate the power of a multi-wavelength approach to probe the merging scenario of such complex and distant systems.}
    
\maketitle
\section{Introduction}\label{sec:introduction}
Mergers of galaxy clusters are unique astrophysical laboratories for observing the assembly of structures and studying the properties of dark matter (DM), the dominant matter component in the Universe. During these events, the fundamentally different properties of galaxies, gas, and DM forming the cluster are dramatically highlighted. Cluster galaxies can be considered as
collisionless particles, affected only by gravity, while the X-ray emitting intracluster medium (ICM) behaves as a fluid slowed down by hydrodynamical processes. 
If DM is collisionless, it should behave like the galaxies, and therefore could be separated from the ICM during particular phases of mergers. Such offsets have indeed been observed in a few galaxy clusters (\citealt{clowe2004,clowe2006}, \citealt{bradac2008}, \citealt{merten2011}, \citealt{dawson2012}, \citealt{gastaldello2014}, \citealt{harvey2015}, \citealt{wittman2023}). The first object where such offset was detected is 1E0657-56; the famous Bullet Cluster \citep{clowe2004,clowe2006}. The X-ray image of this cluster revealed a bullet-like sub-cluster, exiting the core of the main cluster, preceded by a prominent shock front and by a galaxy
density clump, associated with a clear DM concentration traced by the weak lensing (WL) map of the
gravitational potential \citep{clowe2004}. This observation is often considered as a smoking gun for the presence of DM, since it ruled out alternative models of modified gravity \citep{clowe2006}. Furthermore, it allowed one for the first time to use the observation of an offset between DM and baryonic components to provide astrophysical constraints on the self-interaction cross-section of DM particles (\citealt{markevitch2004}, \citealt{randall2008}). 
Such achievements are only possible thanks to the exquisite quality of \chandra\ and \textit{Hubble} Space Telescope (HST) observations, and thanks to the relatively simple geometry and favourable line of sight, which have allowed astronomers to understand and reconstruct the history and dynamics of the merger. Indeed, a good understanding of the merger history of a cluster is a crucial factor for performing a reliable association between gas and galactic substructures and mass clumps observed in the WL maps. Ideal candidates for these studies are unambiguous bullet-like systems, in a merging phase just after the pericentre. Unfortunately, major merger events observed shortly after the first core passage for maximal separation of the components happening on the plane of the sky are rare in the cluster population \citep{shan2010}. 

\begin{figure}
\centering
\resizebox{1\columnwidth}{!}{\includegraphics[trim=60 380 0 50]{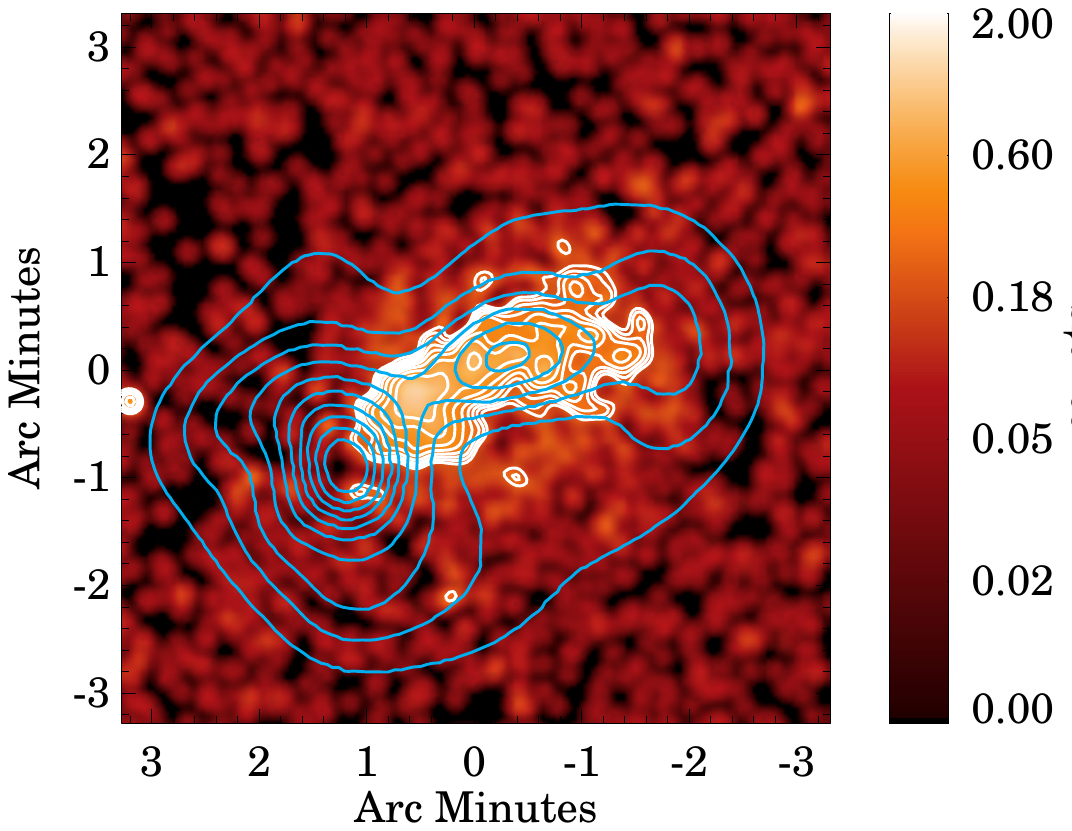}} 
\caption{Image of \amas\ obtained with the first \chandra\ snapshot observation, smoothed with a Gaussian with a full width at half maximum of 6 arcsec. White contours mark the X-ray isophotes, while cyan contours represent the density levels of candidate cluster galaxies within 1 Mpc from the X-ray peak, obtained with archival WFI photometric data. The image is centred on the optical cluster centre reported in Table \ref{tab:OptProp}.}
\label{fig:oldobs}
\end{figure}

In the last decade, the search for new massive galaxy clusters has benefited from the results of several cluster surveys based on the Sunyaev-Zeldovich (SZ, \citealt{sz1972}) effect, which provided catalogues containing thousand of clusters such as those of the Atacama Cosmology Telescope \citep{hamilton2021}, the South Pole Telescope (\citealt{bleem2020}), and the \textit{Planck} full sky catalogue (\citealt{planckesz}, \citealt{planck_psz2}). Since SZ surveys are less biased than X-ray surveys towards relaxed objects (e.g. \citealt{eckert2011,rossetti2016,rossetti2017}), they provided many new merging cluster candidates. We recently performed a complete analysis of the 24 most massive clusters in the PSZ2 catalogue \citep{planck_psz2} with \chandra\ data at $z>0.5$. Most of them ($75\%$) are classified as unrelaxed with morphological indicators derived from X-ray images. Among them, we were particularly impressed by the X-ray image of \amas, a massive (M$_{500}$\footnote{M$_{\Delta}$ is the total mass content within R$_{\Delta}$, which is defined as the radius of the sphere inside which the average density is $\Delta$ times the critical density of the Universe at the cluster redshift.}$ \sim 7.8 \times 10^{14} M_{\odot}$) and high-redshift ($z_{phot} \sim$ 0.66) cluster detected with $S/N=6.6$ in the PSZ2 catalogue  \citep{planck_psz2}. 
Although obtained with a short \chandra\ observation of $\sim$ 15 ks, the image that we report in Fig. \ref{fig:oldobs} highlights a comet-like morphology with a bright circular region and a tail elongated along the north-west (NW) sector similar to the morphology of ‘El Gordo’ (\citealt{menanteau2012}, \citealt{jee2014}, \citealt{caminha23}). We analysed the optical data available in the public archives (ESO 2.2m Wide Field Imager (WFI)  images in three bands)  and produced a galaxy density map (contours shown in Fig.  \ref{fig:oldobs}), which shows two clear peaks in the galaxy distribution, aligned with the X-ray tail. Furthermore, the peak of the X-ray emission is not associated with a galaxy concentration but lies between the two main peaks of the galaxies' distribution. The SE galaxy clump precedes the gas peak in the approximate direction of the merger suggested by the comet-like
shape of the X-ray emission. Assuming that the DM distribution follows the galaxy distribution and its two
peaks, \amas\ could be considered as a candidate analogue of the Bullet Cluster, caught in a similar merger phase. Moreover, the projected separation between gas and galaxies in the south-east (SE) peak is 55 arcsec, corresponding to $\sim 350$ kpc at the cluster redshift, one of the largest observed so far.
Therefore, \amas, dubbed ‘Planck bullet’, is likely an extreme system, possibly
undergoing a major bimodal merger, observed shortly after core passage, featuring an extreme separation
between gas and galaxies. It is thus a very promising candidate for studies of DM-baryon separation, both
as a single object and in combination with a few well-characterised merging clusters in a similar phase. For these reasons, we started a multi-wavelength observational campaign of \amas, with deeper or new observations in the X-ray, optical, and radio bands. Our analysis reveals the presence of an additional substructure interacting with \amas\ and a complex temperature spatial distribution that makes the interpretation of the merging scenario of this cluster particularly puzzling.
In this paper, we present our results based on the new \chandra\ data and spectroscopic measurements of the galaxy population with Telescopio Nazionale Galileo (TNG). \ccob{We shall present in separate forthcoming papers the results of the lensing analysis with HST data and the radio observations performed with the upgraded Giant Metrewave Radio Telescope (uGMRT).}

The paper is organised as follows. In Sect. \ref{sec:section2}, we present the X-ray analysis and results, in Sect. \ref{sec:section3} we present the optical analysis of the photometry and spectroscopy of the member galaxies, in Sect. \ref{sec:section4} we present our interpretation of the merging scenario and the measurement of the cluster total mass combining X-ray and optical datasets, and in Sect. \ref{sec:section5} we draw our conclusions.  We shall present in forthcoming papers the mass distribution of \amas, reconstructed through gravitational lensing on HST data, and the uGMRT radio observations, showing the presence of a radio relic.

We adopted a flat $\Lambda$-cold DM cosmology with $\Omega_{\mathrm{M}} = 0.3$, $\Omega_\Lambda = 0.7$, and H$_{0} = 70$ km/s/Mpc. In the assumed cosmology, 1\arcm corresponds to $\sim 387$ \kpc at the cluster redshift ($z=0.556$, Sec. \ref{sec:OptProp}). The galaxy velocities derived in this work are line-of-sight velocities (los-velocities) determined from the redshift, $\mathrm{V=cz}$.
All the fits were performed via $\chi^2$ minimisation and errors are reported with a confidence level (c.l.) of 68\% unless specified otherwise.

\section{Intracluster medium observations and properties}\label{sec:section2}
\subsection{\chandra\ X-ray data preparation}\label{sec:data_section}
The cluster \amas\ has been observed by \chandra\ using the Advanced CCD imaging spectrometer-Imaging (ACIS-I, \citealt{garmire2003}) for a total of 161 ks divided between eight observations. The observation IDs are 18295, 23852, 24346, 24347, 24348, 24996, 26368, and 26369.

The \chandra\ observations were processed following the procedures described in Sect. A.1 of \citet{bartalucci2017}. Briefly, we used the Chandra Interactive Analysis of Observations (CIAO, \citealt{fruscione2006}) tools version 4.15 and the Chandra ACIS calibration database version 4.10.2 as of November 2022. We applied the latest calibration files and mapped the bad pixels using the chandrarepro tool. Particle contamination was reduced by applying the standard grade selection and very faint mode filtering.\footnote{\url{cxc.cfa.harvard.edu/cal/Acis/Cal_prods/vfbkgrnd/}}

Observing periods affected by flares were removed following the procedures described in \citet{markevitch2006} and in the Chandra background COOKBOOK. The light curve was extracted from the four ACIS-I CCDs with a binning of 259 seconds in the [0.3-12] keV band. We removed from the analysis the time periods in which the count rate is greater than $3\sigma$ with respect to the mean count rate.
The effective exposure time after the cleaning procedures is 145 ks. 
We merged the datasets of the eight observations after the cleaning procedures described above. Images and profiles presented in this work were extracted from the merged dataset unless stated otherwise. The vignetting was corrected using the weighting scheme described in Appendix B of \citet{bartalucci2017}.

We identified the point sources by running the wavedetect detection algorithm \citep{freeman2002} on exposure-corrected images in the [0.5-1.2] keV, [1.2-2.] keV, and [2-7] keV bands. We used the wavelet scales 1, 2, 4, 8, and 16 in units of pixels. The lists of point sources identified in the three bands were merged and inspected by eye for false detections or missed point sources. \ccob{The regions associated with confirmed point sources were filtered out when producing X-ray surface brightness profiles and spectra. However, the point sources are still shown in images.}

\subsection{Background evaluation}
The components of the X-ray background can be divided into instrumental and astrophysical ones. The former are caused by interactions of high-energy particles with the detectors that are detected as photons. In this work, we evaluated this component by producing mock particle background datasets using the analytical model of \citet{bartalucci14}. These datasets were produced in detector co-ordinates for each observation and we projected them to match the observations and normalised them to the count rate in the [9.5-10.6] keV band \citep{bartalucci14}. As for the observations, these datasets were merged and also arranged in data-cubes. From now on, we refer to these datasets simply as the background datasets.

The sky background is formed by a local component \citep{kuntzsnowden2000} and an extra-galactic component due to the unresolved emission of distant active galactic nuclei (AGN) (for more details, see \citealt{lumb2002,kuntzsnowden2000, giacconi2001}). The sky component can be considered as a uniform emission on the scale considered of $\sim 10'$. The evaluation and subtraction of the sky background procedures are different for the surface brightness profile and spectrum extraction. For this reason, these techniques are described in Sects. \ref{sec:sx_analysis} and \ref{sec:spectrum_analysis} for the former and the latter, respectively.
\begin{figure}[!ht]
\centering
\resizebox{1\columnwidth}{!}{\includegraphics[trim=70 380 0 50]{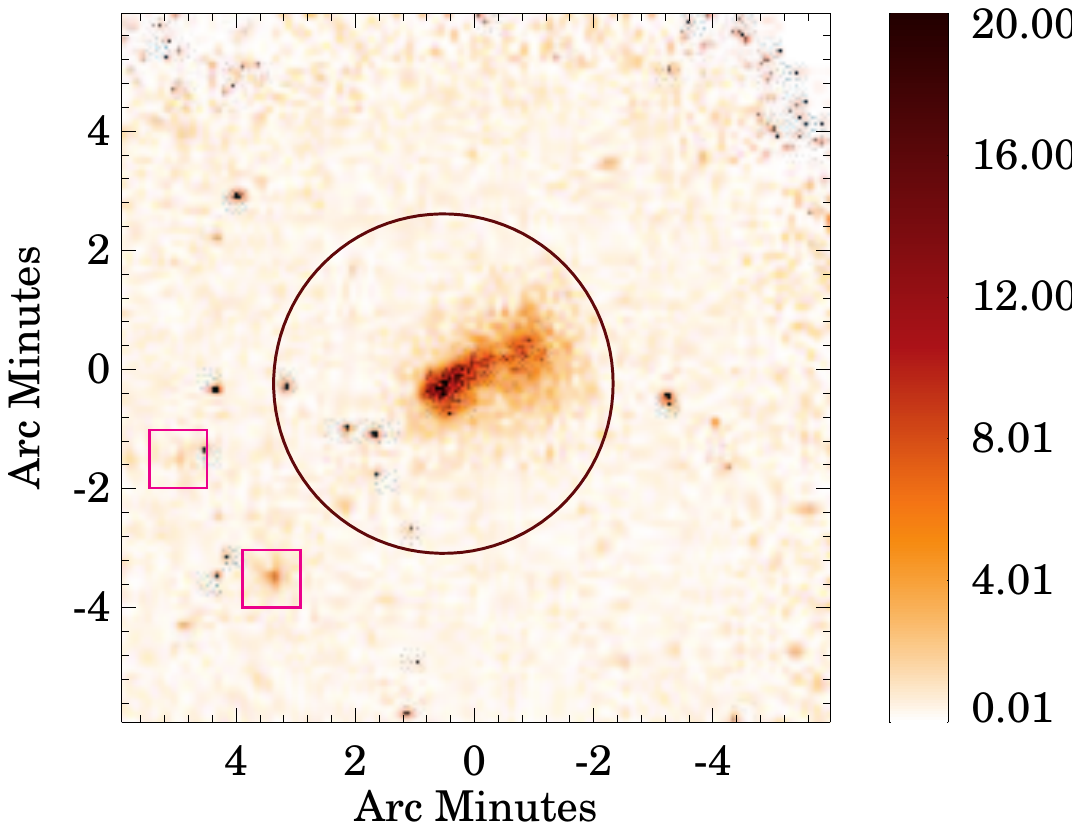}}
\caption{Exposure-corrected and background-subtracted image of \amas\ in the [0.5-2.5] keV band. The images is centred on the X-ray peak. The black circle identifies $\Rvyx$ and the two magenta squares in the eastern sector identify the position of two extended emissions. }
\label{fig:sx_ima}
\end{figure}

\subsection{Imaging}
One of the key point of this work is the imaging analysis of the complex morphology of the cluster. For this reason, we leveraged the imaging tools developed by \citet{bourdin2008} to produce wavelet-cleaned images and temperature maps by arranging the cleaned merged dataset according to the procedure described in \citet{bourdin2008}. This procedure created data-cubes for the observation and the background dataset in which each event is stored according to its energy and position. 

We derived the X-ray image exposure-corrected and instrumental background-subtracted in the [0.5-2.5] keV band shown in \figiac{fig:sx_ima} using these data-cubes. We determined the peak of X-ray emission from this image by applying a  Gaussian smoothing with a kernel $\sigma=5$ pixels corresponding to $\sim 2.5"$. 

\subsection{Surface brightness profile}\label{sec:sx_analysis}
We defined concentric annuli centred on the X-ray peak with a minimum bin of 2". We measured the average vignetting- and exposure-corrected count rate normalised by the area in each annulus in the [0.7-2.5] keV band and subtracted the particle background count rate measured in the same annulus but from the background datasets. 
We evaluated the sky background by identifying the region where the surface brightness profiles flattens; in other words, where the cluster emission is no longer detectable. This annular region is centred on the X-ray peak, the minimum and maximum radii being 3\arcmin\ and 4\arcmin, respectively. We then evaluated the sky background level, measuring the average count rate in this region, and subtracted it from the surface brightness profile. The background-subtracted, exposure- and vignetting-corrected surface brightness profile was then re-binned to have at least $3\sigma$ in each annulus. 

\subsection{Spectrum extraction and analysis}\label{sec:spectrum_analysis}
The X-ray spectra used in this work were extracted using the following procedure. We extracted the photons within the region of interest from the observation dataset. We extracted the particle background events from the same region using the background dataset and subtracted it from the observation spectrum.
We also computed the instrument response files, the Redistribution Matrix File (RMF) and the Ancillary Response File (ARF), using the mkacisrmf and mkacisarf tools of the CIAO suite. The latter was computed at the aim point detailed in \citet{bartalucci2017}.

The sky background was evaluated by extracting the particle background-subtracted spectrum within the same annular region free from cluster emission determined in Sect. \ref{sec:sx_analysis}. We used the sky emission model of \cite{kuntz2008}. It comprises two absorbed thermal emission models, APEC, and one absorbed power law with a fixed slope, $\alpha=1.42$ \citep{lumb2002}. The APEC model parameters are the temperature, metallicity, redshift, and normalisation. Both the APEC metallicity and redshift were fixed to 0.3 and 0, respectively. The absorption of the hydrogen column density along the line of sight was folded in using the photoelectric absorption model phabs, whose values were fixed to  $0.7\times 10^{20}$cm$^{-3}$ and $2.91\times 10^{20}$cm$^{-3}$ for the sky background and cluster components, respectively. The latter value was evaluated using \cite{kalberla2005}. The APEC temperatures and the normalisation of all three components were then determined by fitting the spectrum using the XSPEC \citep{arnaud1996} package and folding the appropriate response files.
The sky background model was then fixed and simply scaled by the ratio of the area of the region of interest to the sky background annulus area. 

We derived the temperature of the ICM following a similar procedure. We fitted the particle background-subtracted spectrum extracted from the region of interest with an absorbed APEC model, accounting for the ICM emission plus the properly re-scaled sky background. 
The redshift was fixed to the value $z=0.556$, determined from optical spectroscopy, as is detailed in Sect. \ref{sec:OptProp}, revising the original estimate provided in the PSZ2 cluster catalogue \citep{PSZ2}, which had been estimated from photometric data. The normalisation, temperature, and abundance were then determined through the fit in the [0.7-12] keV band. All the fitting procedures described above take in account the proper response ARF and RMF files.


\begin{table}
\caption{\footnotesize Global X-ray properties of \amas.}
\label{tab:clus_prop}
\begin{center}
\begin{tabular}{lr}
\toprule
Quantity            & Value     \\
\midrule
RA-DEC primary X-peak  &  $179.50258$; $-10.772888$ \\
RA-DEC secondary X-peak  &  $179.4797357$; $-10.7631486$\\ 
$\Mvyx{}$   & $7.35^{+0.64}_{-0.56} \times 10^{14}$ $[\mathrm{M}_{\odot}]$\\
$\Rvyx$ $^a$    & $1115^{+33}_{-29}$ [kpc]\\
$T_{\textrm X}$ & $6.88^{+0.5}_{-0.44}$ [keV]\\
$M_{gas}(<\Rvyx)$ & $1.24^{+0.10}_{-0.09} \times 10^{14}$ $[M_{\odot}]$\\
\bottomrule
\end{tabular}
\end{center}
\footnotesize{Notes: $^{(a)}$ The radius in arcmin is $\Rvyx=2.85^{+0.08}_{-0.07}$ arcmin. }
\end{table}


\subsection{Cluster redshift}\label{sec:xray_redshift}
Redshift measurements are possible with X-ray data using the position of the ICM emission lines, most notably Iron $K_\alpha$ at $\sim$7 keV for hot clusters. 
We extracted the spectrum from a circular region that maximises the signal-to-noise, that is centred on the primary X-ray peak, and that has a radius of $R=2.05$ arcmin. We applied the fitting procedure described in Sect. \ref{sec:spectrum_analysis}, setting the redshift as a free parameter, and obtained the best-fit value $z_X=0.51^{+0.01}_{-0.02}$. The ACIS-I detector on board \chandra\ suffers from energy resolution degradation.\footnote{\url{https://cxc.harvard.edu/proposer/POG/html/chap6.html}} We argue that at the effective energy of the iron line  uncertainty related to this effect is of the order of $227$ eV.\footnote{We interpolated this value assuming that the degradation is linear as a function of energy between the measurements of Aluminum K$\alpha$ and Manganese K$\alpha$.} This translates into a systematic error of $\delta_{sys}=0.05$. The redshift determined using the X-ray, $z_X=0.51^{+0.01}_{-0.02} \pm 0.05$, is consistent within $1\sigma$ with the redshift determined using optical spectroscopic data presented in Sect. \ref{sec:OptProp}.
\subsection{Intracluster medium distribution}
The ICM spatial distribution of \amas\ is characterised by a comet-like shape elongated in the SE-NW direction, as is shown in \figiac{fig:sx_ima}. The cluster hosts a major bright halo and a tail extending towards the NW. This latter structure is formed by a narrow and elongated emission attached to the main halo and then it enlarges, forming a roundish structure the same size as the main clump. One of the key points for the interpretation of the merging scenario is the significance of the structures identified within the ICM in the X-ray images. For this reason,  we produced an exposure-corrected and background-subtracted X-ray image in the [0.5-2.5] keV band showing the emission detected with a significance greater than $3\sigma$ (see the top right panel of \figiac{fig:wavelet_and_extended}). This map was obtained by applying the procedure described in \citet{bourdin2008}, using wavelet filtering and thresholding B3-spline wavelet coefficients. From now on, we refer to this maps as the wavelet map. The wavelet map confirms the comet-like shape within which we identified a secondary X-ray peak. The co-ordinates of the two X-ray peaks are reported in Table \ref{tab:clus_prop}. The detection of a secondary peak within merging tails could be due to local inhomogeneities or to a complex and patchy gas distribution (e.g. \citealt{eckert17}). An alternative scenario could be that the tail is a secondary halo merging with the main cluster. 
\subsection{Temperature map}

\begin{figure*}[!ht]
\begin{center}
\resizebox{0.9\textwidth}{!}{
\includegraphics[trim=60 360 0 50]{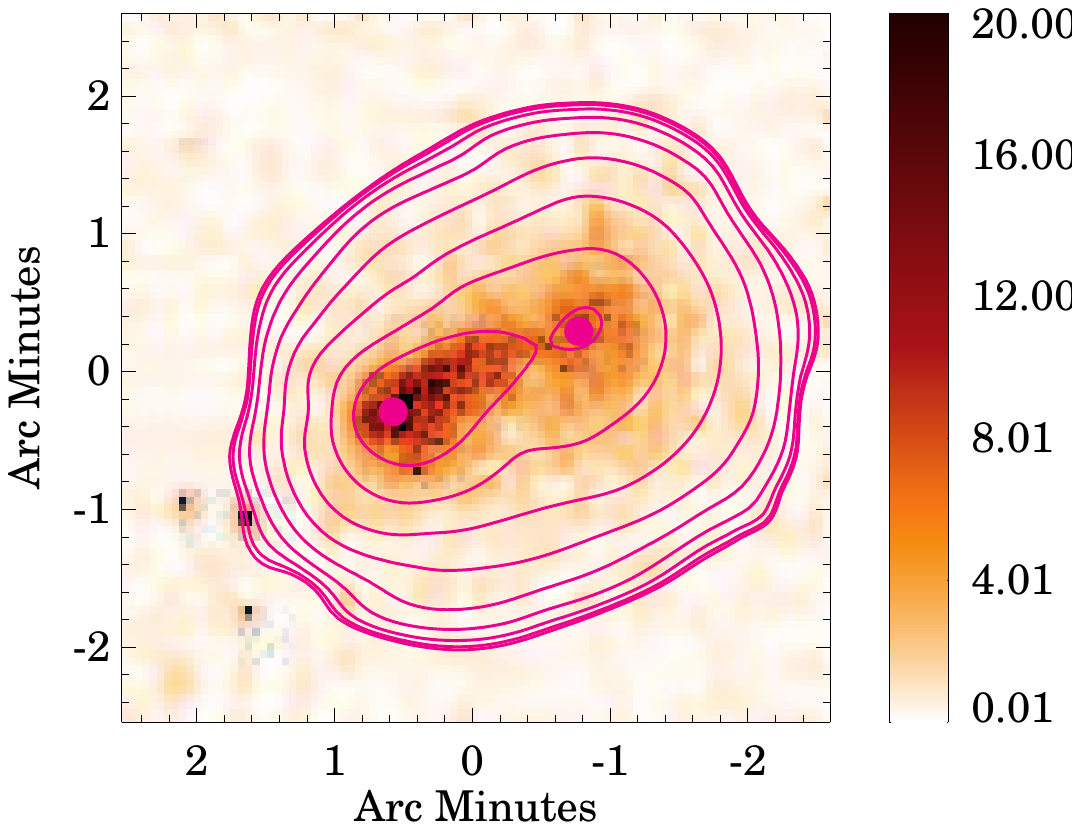}
\includegraphics[trim=60 360 0 50]{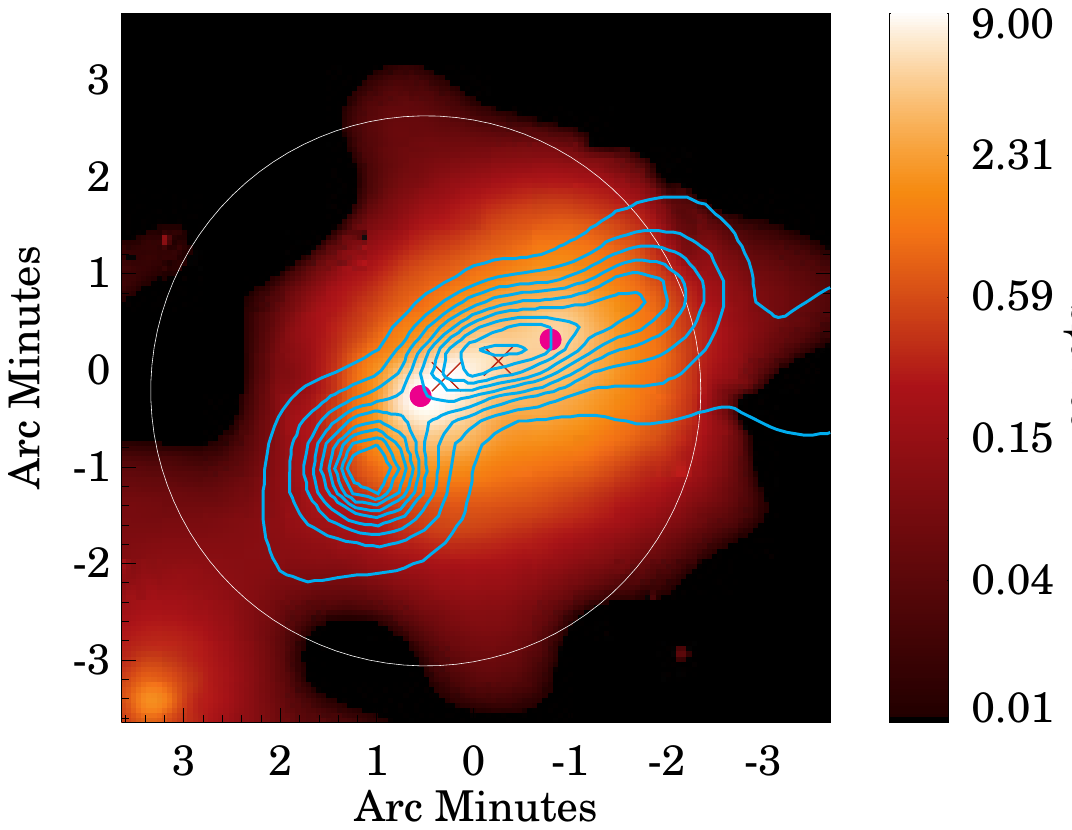}
}

\resizebox{0.9\textwidth}{!}{
\includegraphics[trim=60 360 0 50]{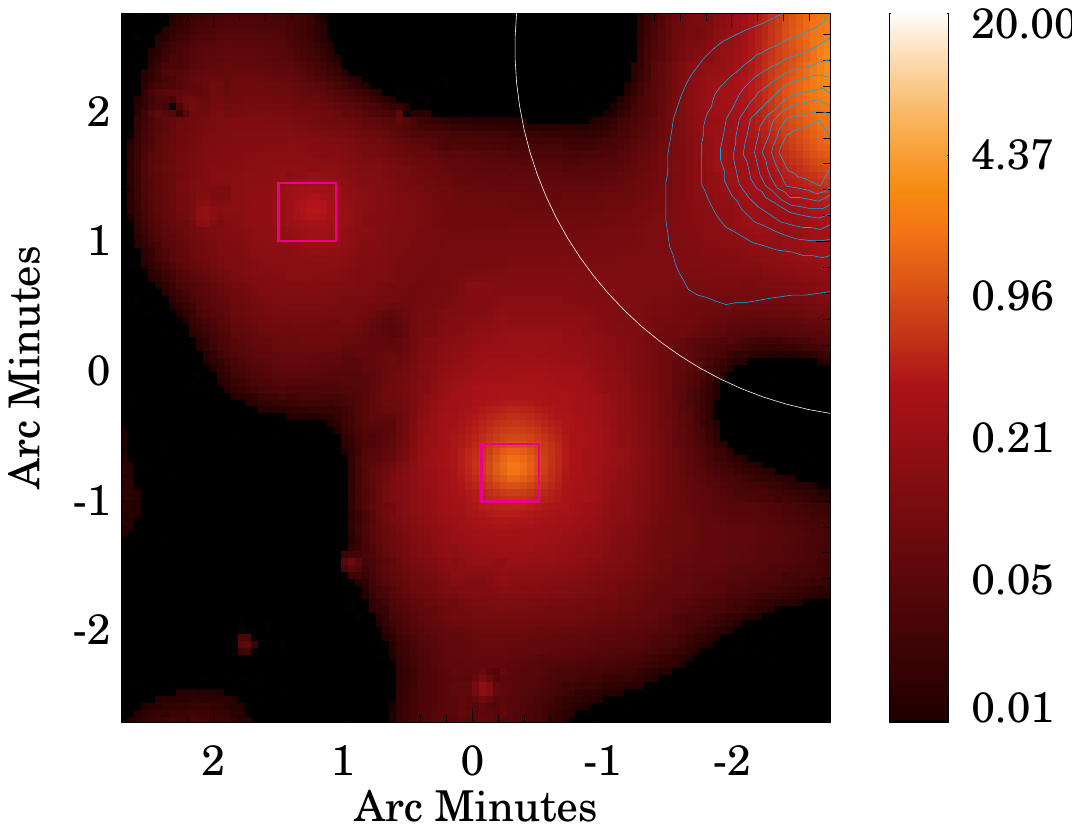}
\includegraphics[trim=60 360 0 50]{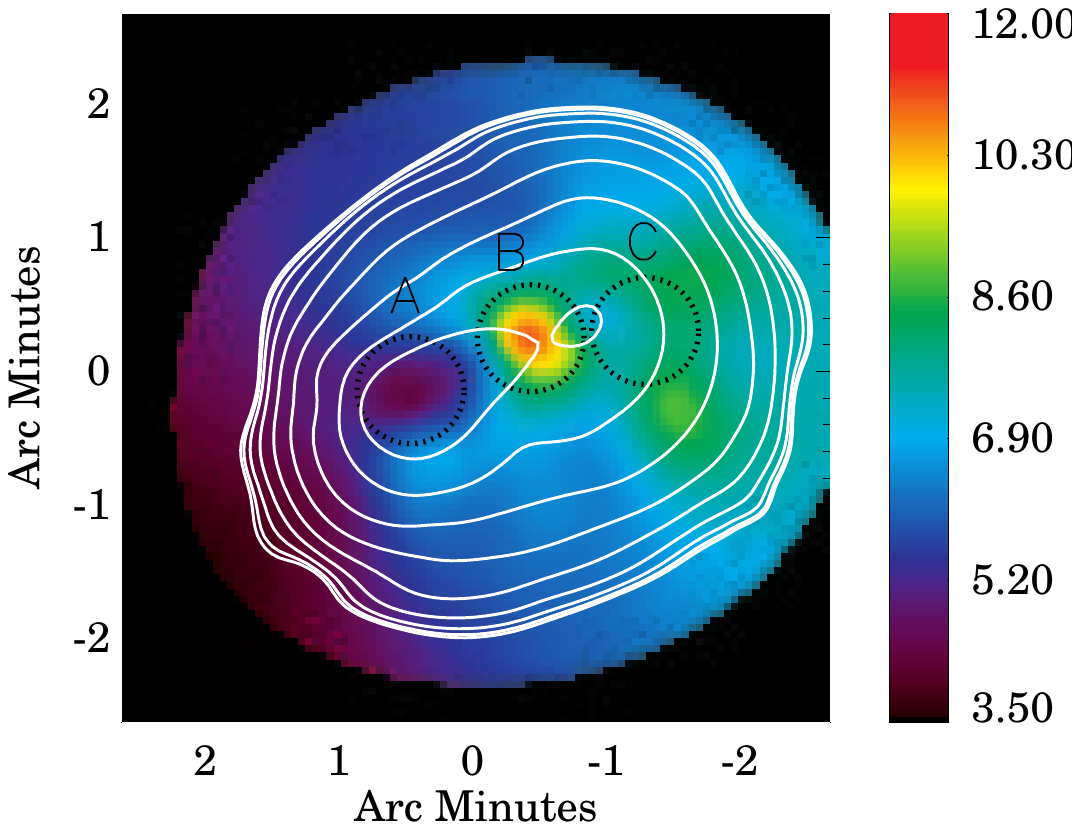}
}
\end{center}
\caption{{Spatially resolved morphological and thermodynamic analysis of \amas. Top left panel: Same as in \figiac{fig:sx_ima}, with the position of the X-ray peaks highlighted by the magenta points and the iso-contours derived from the wavelet-cleaned image in the top right panel. Top right panel: Wavelet-cleaned image in the [0.5-2.5] keV band centred on the central part of the cluster. The white circle encompasses $\Rvyx$. The two crosses identify the position of BCG1 and BCG2 described in Sect. \ref{sec:mem}. The cyan contours identify the iso-levels of the spectroscopic galaxy members of \amas. The map is in unit of counts. Bottom left panel: Offset view of the extended emissions in the E sector highlighted by the magenta squares, as in \figiac{fig:sx_ima}. \ccob{The E emission is interacting with the main cluster, being at the same redshift. The SE is associated with a foreground group emission at z$\sim 0.17$ (see Sect. \ref{sec:2D} for more details).} Bottom right panel: Temperature map of \amas. The white lines identify the iso-contours derived from the wavelet-cleaned map shown in the top right panel. The dotted black circles identify the three regions within which we extracted the spectra to derive the spectroscopic temperature, whose co-ordinates and radius are reported in Table \ref{tab:temperature_check}. The units of the colour bar are keV.}}
\label{fig:wavelet_and_extended}
\end{figure*} 

We derived the spatially resolved temperature map of \amas\ by applying the wavelet filtering method detailed in \citet{bourdin2008} and \citet{bourdin2013}. The details for the application of this method applied to a cluster observed with \chandra\ at similar redshift can also be found in \citet{adam2017}. In particular, for this work we used regions where the minimum size was set to have at least 100 source photons, and the extension of the map was limited to regions where the signal-to-background ratio is greater than 0.3; that is, the extension of the map was limited by the statistics.

The temperature map of \amas\ is shown in the bottom right panel of Fig.~\ref{fig:wavelet_and_extended} with X-ray contours overlaid. The spatial behaviour of the temperature highlights the presence of a roundish hot region, kT$\sim 11$ keV, between the two X-ray halos. We labelled this region B. The coldest region on the map, kT $\sim 4$ keV, corresponds to the central part of the main X-ray halo on the E. We labelled this region A. Finally, the W sector is occupied by a diffuse  hot region, kT $\sim 8$ keV, elongated in the N-S direction, which does not correlate with an X-ray structure or halo. This region was labelled C. 
The co-ordinates of these regions are reported in Table \ref{tab:temperature_check}.

The pixels of the image are highly correlated due to the nature of the algorithm. Furthermore, the resolution of the map is regulated by the statistic as the size of a region with at least 100 counts varies on the image. For this reason, we used this map as qualitative anchor to define three circular regions reported in Table \ref{tab:temperature_check}, from which we extracted the spectrum and measured the temperature following the procedure described in Sect. \ref{sec:spectrum_analysis} to perform a quantitative analysis. The results of the fit for the three regions are reported in Table \ref{tab:temperature_check}. The measured temperatures within the A and B regions confirm the scenario suggested by the temperature map, the temperatures being kT$_{\mathrm{A}}=4.79 \pm 0.28$ keV and kT$_{\mathrm{B}}=9.59^{+2.82}_{-1.31}$ keV. The temperature in the C region, kT$_{\mathrm{C}}=8.48^{+2.34}_{-1.43}$ keV, confirms the temperature map value. However, it is worth noting that the temperatures in the B and C regions are consistent within $1\sigma$. 

\begin{table}
\caption{ {\footnotesize Spectroscopic temperatures of \amas}}\label{tab:temperature_check}
\begin{center}
\begin{tabular}{lcr}
\toprule
Region            & Co-ordinates (RA; DEC)      & Temperature     \\
                  & [degrees, J2000]          & [keV]           \\
\midrule
A               & 179.502; -10.772          & 4.79 $\pm$ 0.28 \\
B               & 179.487; -10.765          & $9.59^{+2.82}_{-1.31}$ \\
C               & 179.472; -10.764          & $8.48^{+2.34}_{-1.43}$ \\
\bottomrule
\end{tabular}
\end{center}
\footnotesize{Notes: Each region corresponds to a circle centred on the listed co-ordinates, the radius of which is 0.4 arcmin.}
\end{table}
\begin{figure*}[!ht]
\begin{center}
\resizebox{1\textwidth}{!}{
\includegraphics[]{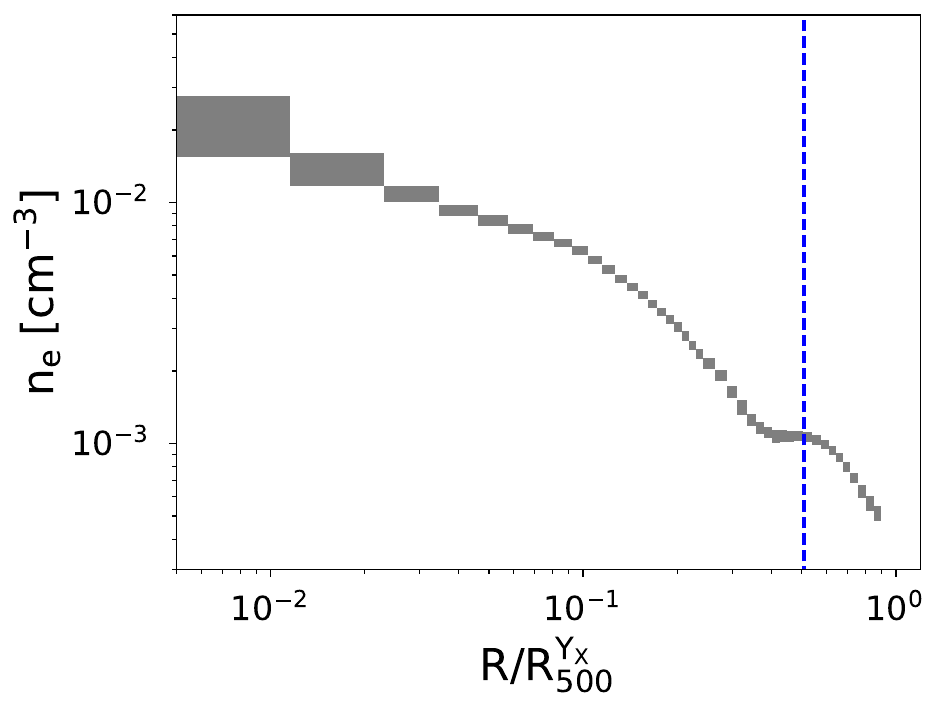}
\includegraphics[]{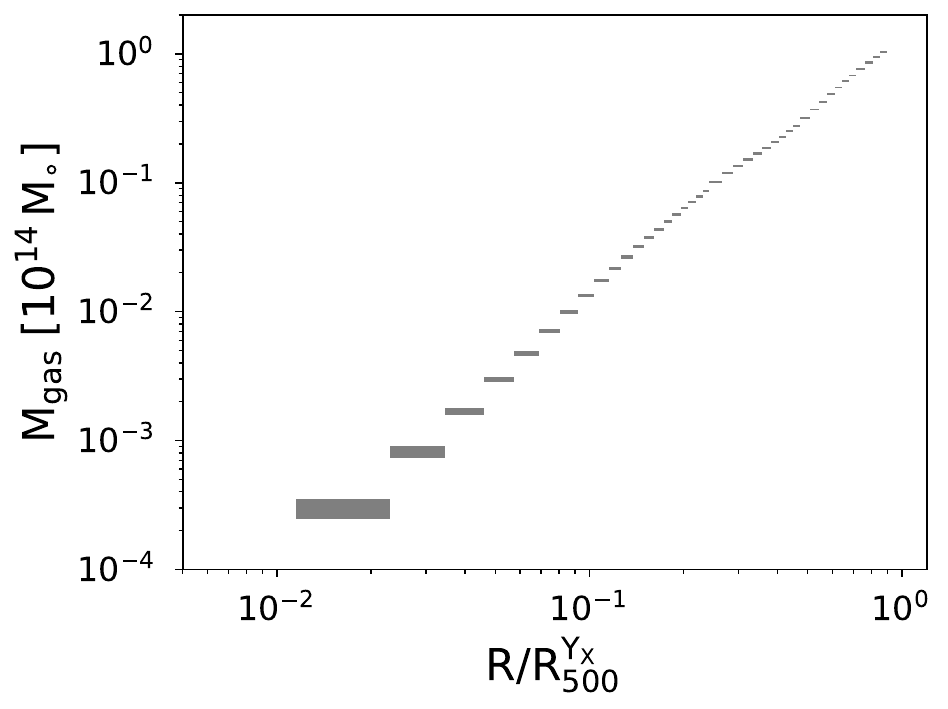}
}
\end{center}
\caption{{Thermodynamic radial profiles of \amas\ scaled by $\Rvyx$. Left panel: De-projected density profile centred on the X-ray peak. The vertical dotted blue line marks the distance between the X-ray peak and the emission of the peak of the tail to the W.  Right panel: Same as the left panel except for the fact that we show the gas mass profile. The width of the envelope along the y axis represents the $1\sigma$ uncertainty in both panels.}}
\label{fig:thermo_profiles}
\end{figure*} 
\subsection{X-ray mass}\label{section:xray_mass}
\label{sec:x-mass}
Starting from the \chandra\ data, we derived the cluster mass within R$_{500}$, $\Mv$, from the relation between the cluster mass and the proxy, $\mathrm{Y_{X}}$. This proxy was introduced by \citet{krav2006} and is defined as the product of the temperature measured in the [0.15-0.75]R$_{500}$ region, T$_{\mathrm{X}}$,  and the gas mass measured within R$_{500}$, M$_{\mathrm{gas,500}}$. \ccob{The relation M$_{500} - \mathrm{Y_{X}}$ was calibrated by \citet{arnaud2010} using the hydrostatic mass of local galaxy clusters at $z \sim 0.2$.}
It is argued that this relation is predicted to have the minimum intrinsic scatter and its evolution with redshift is within the $1\sigma$ uncertainties, both in terms of shape and normalisation, as is discussed in \citet{lebrun2017}.

To measure the gas mass, we derived the ICM radial density profile, de-projecting the surface brightness profile using the technique detailed in \citet{bartalucci2017} (see \citealt{croston2006} for the de-projection with regularisation technique). The scaled density profile centred on the main X-ray peak is shown in the left panel of \figiac{fig:thermo_profiles} and its shape is affected by the particular geometry of the cluster. The density profile flattens in the radial range of R=[$0.4,0.6] \Rvyx$, corresponding to the roundish structure within the tail, the vertical line identifying the position of the secondary peak. The size of this feature is $\sim 0.2\Rvyx$, which corresponds to $\sim 0.6$ arcmin ($\sim 220$ kpc), while the extension of the tail is almost $1.5$\arcmin. For this reason, we argue that the feature in the density profile confirms the scenario in which the tail is formed by a secondary halo surrounded by diffuse gas due to the interaction with the main halo.

The right panel of \figiac{fig:thermo_profiles} shows the gas mass radial profile obtained by integrating the density within concentric shells. The flat feature is not visible in the gas mass profile because it is an integrated quantity.

With the spatially resolved gas mass profile, we were able to derive iteratively the cluster mass within R$_{500}$. We first guessed the value of $\Rvyx$ and derived the temperature within the [0.15-0.75]$\Rvyx$ region following the procedure described in Sect. \ref{sec:spectrum_analysis}. We computed the gas mass within that radius to compute $\mathrm{Y_{X}}$ and used the M$_{500} - \mathrm{Y_{X}}$ relation to derive the value of $\Mvyx$. From this latter value, we derived $\Rvyx$ and compared it with the initial value. This procedure was repeated until the difference between the initial and derived $\Rvyx$ was less than $1\%$. We found that the mass enclosed within $\Rvyx=1115^{+32}_{-29}$ kpc, corresponding to $\Rvyx=2.85^{+0.08}_{-0.07}$ arcmin, is $\Mvyx=7.35^{+0.64}_{-0.56} \; 10^{14} \mathrm{M}_{\odot}$. This is consistent with our revised estimate of the SZ mass from Planck data, assuming an optical redshift of M$_{500}^{SZ}=(7.4\pm0.6) \times 10^{14}$M$_\odot$. 
These values, together with other parameters obtained by the fit, are reported in Table \ref{tab:clus_prop}.

\subsection{Extended X-ray sources}\label{sec:extended_sources}
We detected two significant diffuse emissions in the east sector that are highlighted with magenta squares visible in the count and the wavelet-cleaned map in \figiac{fig:sx_ima} and in the bottom left panel of \figiac{fig:wavelet_and_extended}, respectively. Using PanSTARRS photometry of the most prominent galaxy coincident (within a few arcsecs) with the X-ray peak of each source, we estimate a
redshift of z=$0.19 \pm 0.03$ and z= $0.57 \pm 0.04$ for the SE and east-south-east (ESE) sources, respectively.

The SE source is not associated with the main cluster and just happens to be a foreground group. Its X-ray peak emission co-ordinates are  RA, DEC=[179.551,-10.827] deg. We measured a luminosity of L =($4 \pm 0.5) \times 10^{42}$ erg/s for this group and by using the L--M relation of \cite{lovisari2015} we estimated the mass to be M$_{500}=(3 \pm 1) \times 10^{13}$M$_{\odot}$.

The photometric redshift of the ESE extended source is compatible within $1\sigma$ with the main cluster redshift, 
suggesting that this is an interacting or just in-falling galaxy group. Its X-ray peak emission co-ordinates are  RA, DEC=[179.577,-10.793] deg. Unfortunately, the statistic of the X-ray data for that object is not sufficient to infer other properties.

\section{Optical data of galaxies, analysis, and results}\label{sec:section3}

\subsection{New optical  data and redshift catalogue}\label{sec:tng}

We obtained photometric data of the field of \amas\ with the instrument DOLoRes\footnote{\url{www.tng.iac.es/instruments/lrs}} of the Italian Telescopio Nazionale {\em Galileo} (TNG) in January 2020. We took eight images in both the SDSS $r$ and in the $i$ band with T$_{\exp}$ = 180 s, in photometric conditions and a stable seeing of 0.8\arcss--0.9\arcss. The images were reduced in a standard way, by subtracting the bias and dividing for a master flat-field frame. We computed precise astrometric solutions, using as a reference the Gaia DR3 star catalogue (e.g. \citealt{lindegren2021}) using IRAF\footnote{IRAF is distributed by the National Optical Astronomy Observatories, which are operated by the Association of Universities for Research in Astronomy, Inc., under cooperative agreement with the National Science Foundation.} tasks. We produced a co-added image for each filter and identified galaxies using the SExtractor package \cite{bertin1996}. We were able to estimate the $r$ and $i$ magnitudes and $r-i$ colours for 1925 galaxies in a field of view of $\sim$8.6\arcmm$\times 8.6$\arcmm. The photometric
calibration was performed using PanSTARRS $r$ and $i$ magnitudes of stars in the observed field, taking into account the transformation equations between the PanSTARRS and SDSS magnitude systems. We estimate that our photometric catalogue is complete down to $r\sim 23.5$.

We performed multi-object spectroscopic (MOS) observations of \amas\ at the TNG in March and April 2021 in the framework of the program A43TAC$_{-}$1 (PI: W. Boschin). We used DOLoRes with the grism LR-R to observe 77 galaxies with two MOS masks. In particular, the choice of target galaxies was based on a colour and magnitude selection. The total exposure time was 9 ks for both masks. Spectral reduction and radial velocity estimation were performed using the standard IRAF tasks and the cross-correlation technique \cite{tonry1979}. We obtained velocity estimates for 66 galaxies. The redshifts of three additional galaxies, IDs 3, 54, and 86 in  Table~\ref{catalogG282a}, were estimated by measuring the wavelength locations of outstanding emission lines in their spectra. The median value of the uncertainties in the velocity measurements is 107 km s$^{-1}$.
We also used archival spectroscopic data of \amas\ taken with the instrument
FORS2\footnote{\url{www.eso.org/sci/facilities/paranal/instruments/fors.html}} of the ESO Very Large Telescope (VLT). These data consist of one MOS mask observed in January 2014 with the grism 300I and a total exposure
time of 5.24 ks. We reduced the data in the manner described above and obtained velocity estimates for 23 galaxies, one of them observed also by the TNG. The median value of the velocity errors for galaxies measured with FORS2 data is 73 km s$^{-1}$.
The velocity catalogue for the 91 galaxies is reported in Table \ref{catalogG282a} in Appendix \ref{appendix} (see also Fig.~\ref{fig:optical_image}). The $r$ and $i$ magnitudes are available for all but one galaxy.

\subsection{Cluster member selection}\label{sec:mem}

\begin{figure}
\centering
\resizebox{\hsize}{!}{\includegraphics[trim=0 150 0 120]{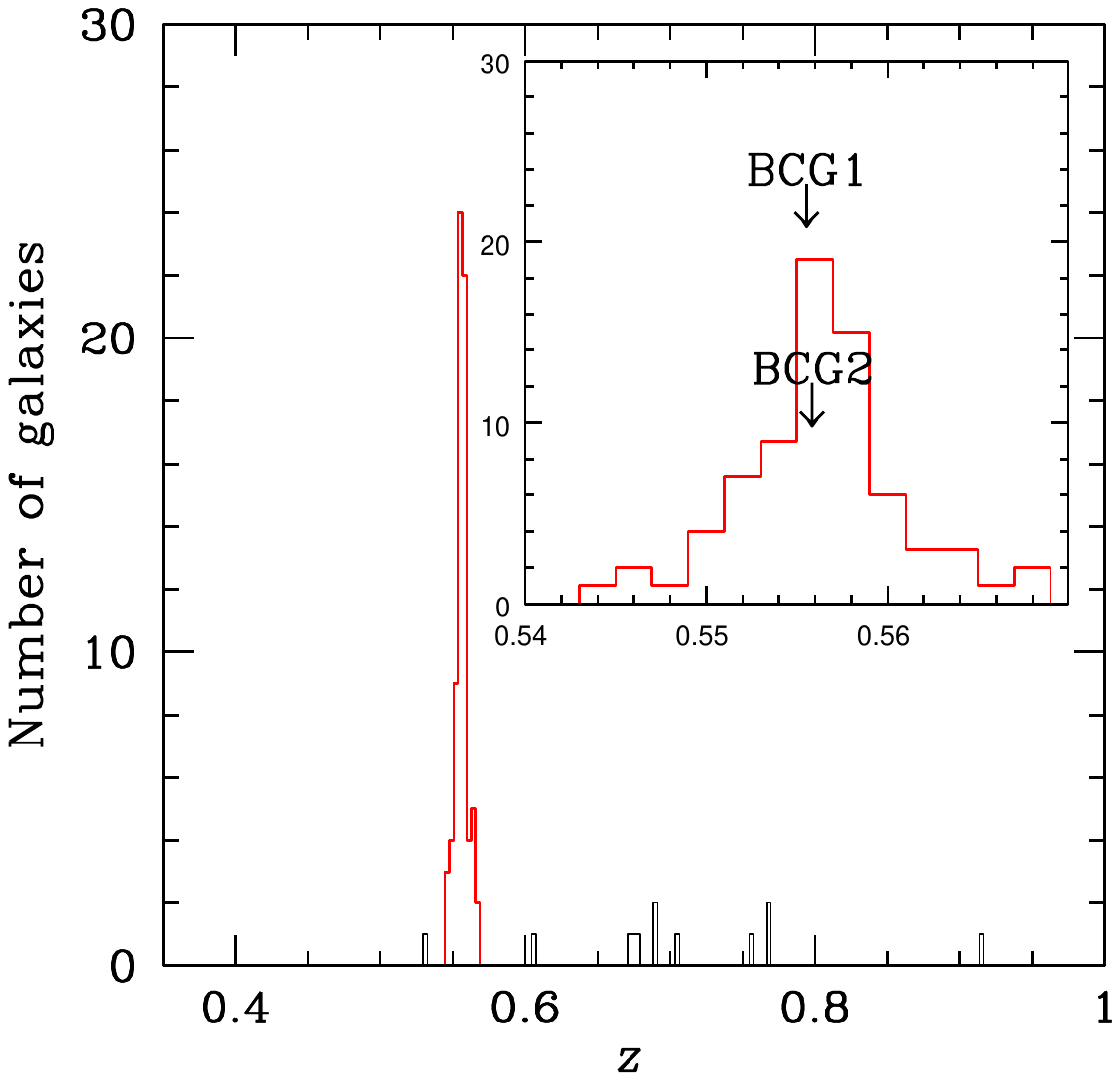}}
\caption
{Histogram referring to
  all the galaxies with a spectroscopic redshift in the region
  of \amas.  The heavy, red line histogram refers to the 73 galaxies
  assigned to the G282 peak according to the 1D-DEDICA reconstruction
  method.  The inset figure shows the 73 member galaxies and the arrows indicate the redshift of BCG1 and BCG2.}
\label{fighisto}
\end{figure}

We applied the two-step method known as ‘peak+gap’ (P+G) already applied by \cite{girardi2015} to select cluster members among the 91 galaxies of our spectroscopic catalogue. The first step is the application of the 1D adaptive-kernel method (1D-DEDICA; \citealt{pisani1993,pisani1996}; see also \citealt{girardi1996}). This procedure detects \amas\ as a peak at $z\sim0.556$ populated by 73 galaxies, as is shown in  
Fig.~\ref{fighisto}.

\begin{figure}
\centering
\resizebox{\hsize}{!}{\includegraphics[trim=0 150 0 150]{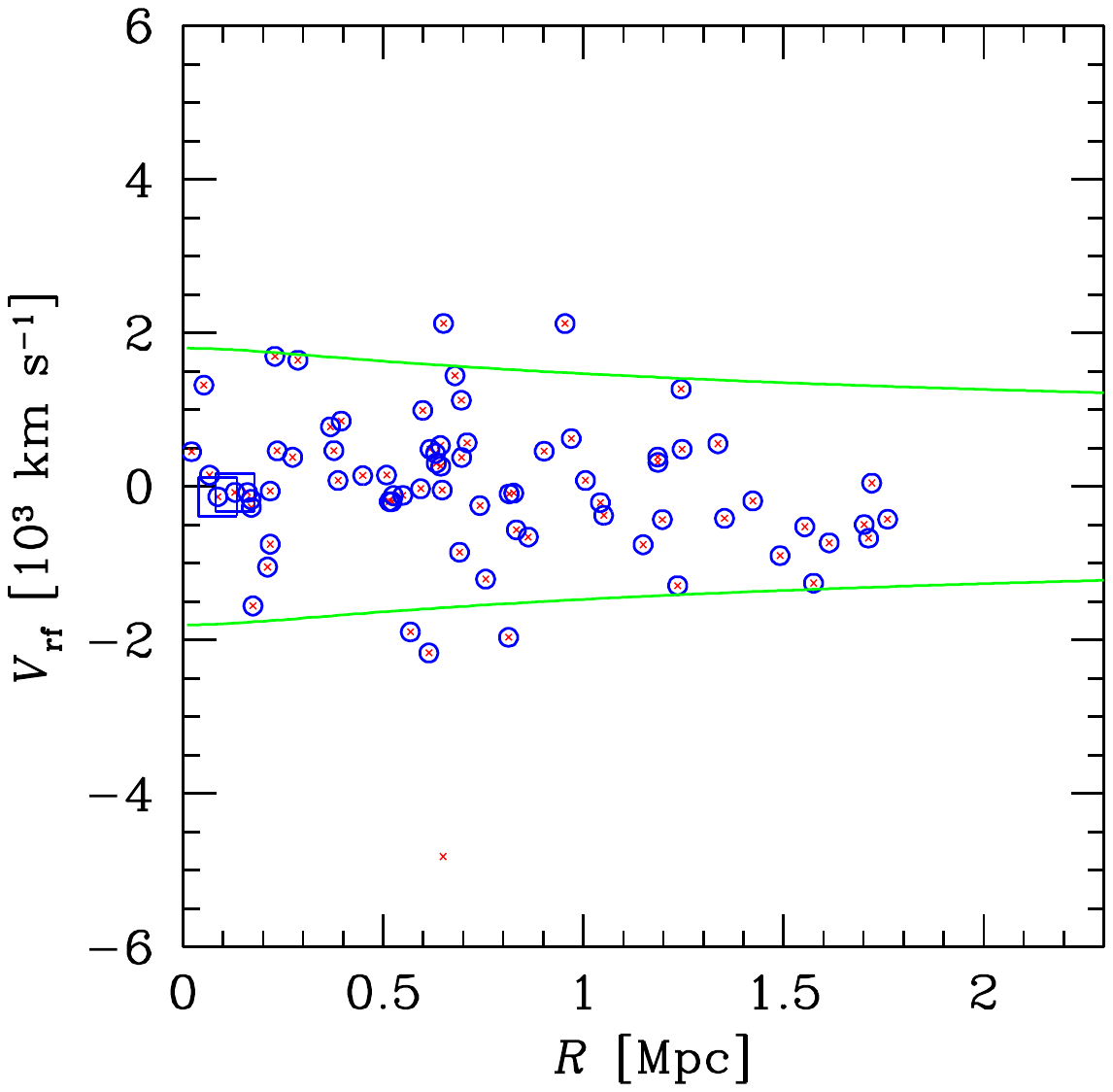}}
\caption
{Rest-frame velocity, $V_{\mathrm{rf}} = (V-<V>)/(1+z)$, vs projected
  clustercentric distance, $R$, for galaxies that have redshifts within 6000
  \ks of the mean cluster velocity (small red crosses).  Blue
  circles indicate cluster members. Large blue squares point out to
  BCG1 and BCG2. Green curves contain the region where $|V_{\mathrm{rf}}|$
  is smaller than the escape velocity (see text).}
\label{figvd}
\end{figure}

In a second step, we combined the space and velocity information in the ‘shifting gapper’ procedure (\citealt{fadda1996}, \citealt{girardi1996}). This procedure excludes galaxies that are too far away in velocity from the main body of galaxies within an annulus around the centre of the system; in other words, ones that are farther
away than a fixed velocity distance called the velocity gap. The position of the annulus is shifted with increasing distance from the centre of the cluster. The procedure is repeated until the number of
cluster members converges on a stable value. Fadda et al. (\citeyear{fadda1996}) suggested a velocity gap of $1000$ \ks in the cluster rest frame and an annulus size of 0.6 \h or more to include at least 15 galaxies.  
We confirmed all the 73 candidate cluster members by following the procedure described above. Figure~\ref{figvd} shows galaxies in the projected
phase-space. \ccob{For the sole purpose of highlighting the cluster member region, we also plot the escape velocity curves computed assuming a NFW mass density profile, according to the recipe of den Hartog \& Katgert (\citeyear{hartog1996}) and using the mass estimated in Sect. \ref{sec:3D}.}

\begin{figure*}[!ht]
\begin{center}
\resizebox{1\textwidth}{!}{
\includegraphics[width=\textwidth,angle=270,trim=0 0 0 0]{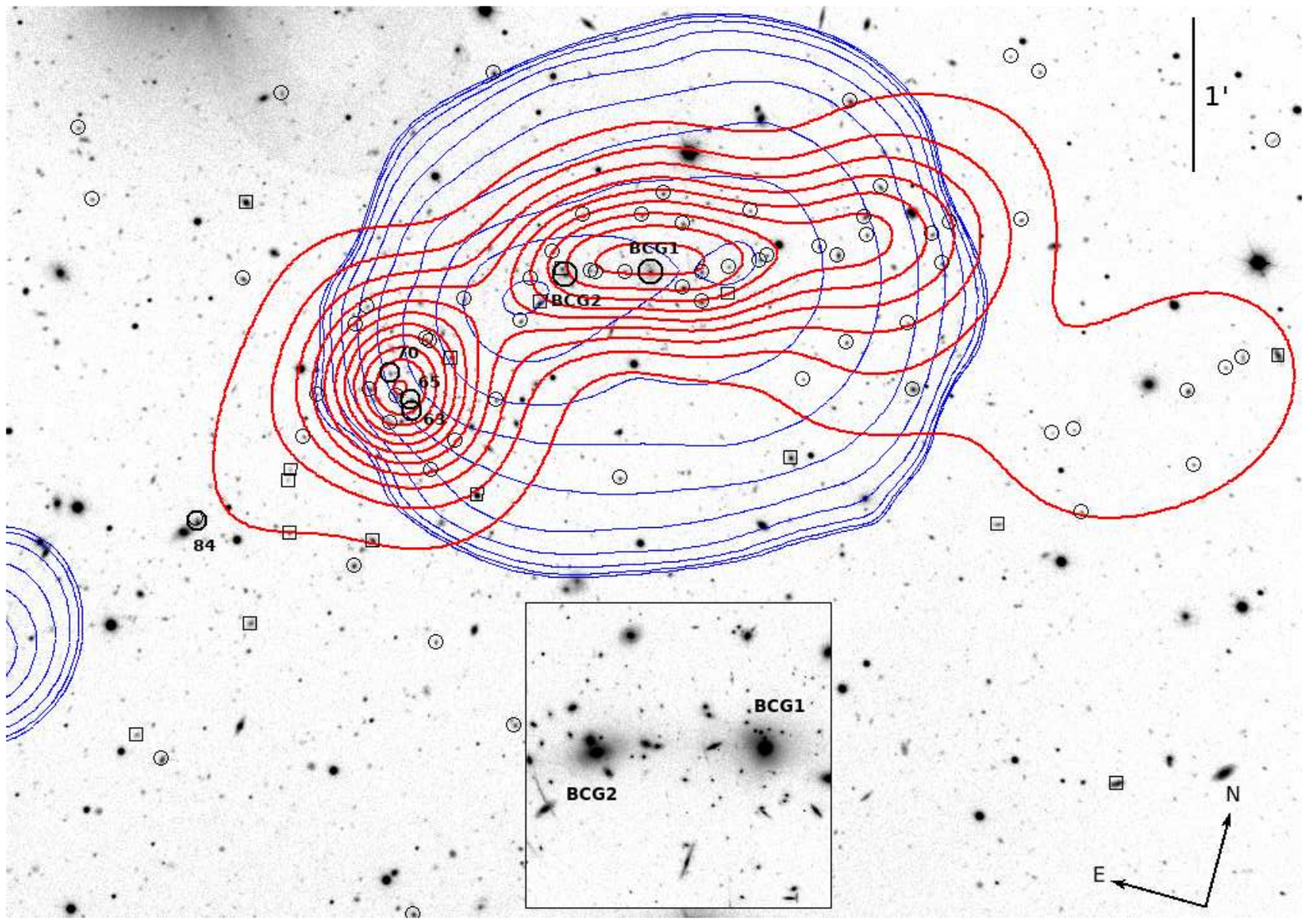}
}
\end{center}
\caption{{TNG $r$-band image of \amas. Labels
  indicate the two brightest cluster members and other galaxies
  mentioned in the text. The blue lines identify the X-ray iso-contours derived from the wavelet-cleaned map shown in the top right panel of Fig.~\ref{fig:wavelet_and_extended}. The red contours represent the isodensity levels obtained with the 2D-DEDICA method applied to the sample of spectroscopic members. Black circles and boxes highlight the member and non-member galaxies, respectively. The inset box on the bottom is a zoom on the central region of the cluster highlighting BCG1 and BCG2 (HST data).}}
\label{fig:optical_image}
\end{figure*}

The  spatial distribution of 73 member galaxies is shown in \figiac{fig:optical_image}.
The central region of \amas\ is dominated by two prominent galaxies. The galaxy ID~39 in Table~\ref{catalogG282a} is the brightest cluster galaxy, $r = 19.58$ (hereafter BCG1). We identified another bright galaxy 34\arcs in the east direction from BCG1, reported as ID~50 in Table~\ref{catalogG282a} with $r = 19.99$ (hereafter BCG2). 
BCG2 is elongated in the ESE-NWN direction, as the central part of the cluster is (see \figiac{fig:optical_image}), while BCG1 seems elongated in the perpendicular direction
(see Fig.~\ref{fig:optical_image}, insert).
In this section, we adopt the centre of \amas\ by computing a flux-weighted average between the two BCG positions: the co-ordinates are reported in Table~\ref{tab:OptProp}. 

\subsection{Optical cluster properties}\label{sec:OptProp}

The analysis of the velocity distribution of the 73 cluster members was performed using the biweight estimators for location and scale included in ROSTAT statistical routines of \cite{beers1990}. The mean redshift of the cluster is $\left<z\right> = 0.5566\pm0.0003$, corresponding to $\left<V\right> = 166\,759\pm98$ \kss.  We estimated the velocity dispersion, $\sigma_{\mathrm{V}}$, by applying the cosmological correction and the standard correction for velocity errors \cite{danese1980}. We obtained $\sigma_{\mathrm{V}} = 833_{-89}^{+103}$ \kss, the errors being estimated using a bootstrap technique. 
These properties are summarised in Table \ref{tab:OptProp}.

The velocity distribution was analysed for possible deviations from
the Gaussian distribution. We calculated the shape estimators proposed
by Bird \& Beers (\citeyear{bird1993}; see their Table~2); that is, the skewness, the
kurtosis, the tail index, and the asymmetry index. We found no evidence of possible deviations from the Gaussian distribution. According to the Indicator test \cite{gebhardt1991}, neither BCG1 and BCG2 have a peculiar velocity (see also Fig.~\ref{fighisto}, inset).

We derived the cluster mass using as a proxy the velocity dispersion and computed the mass $M_{200}$ within $R_{200}$ using the theoretical relation
between $M_{200}$ and the velocity dispersion reported in  Eq.~1 of \cite{munari2013}, where it is verified with clusters extracted from numerical simulations..  We derived $M_{200,\mathrm{opt}} = 4.7_{-2.0}^{+2.2}\times 10^{14}M_\odot$ within $R_{200,\mathrm{opt}} = 1.3_{-0.14}^{+0.16}$ Mpc. The uncertainties for $R_{200,\mathrm{opt}}$ and $M_{200,\mathrm{opt}}$ were calculated using the error propagation of $\sigma_{\mathrm{V},200}$ ($R_{200,\mathrm{opt}}\propto \sigma_{\mathrm{V},200}$ and $M_{200,\mathrm{opt}}\propto \sigma_{\mathrm{V},200}^3$) and an additional uncertainty of $10\%$ for the mass due to the scatter shown in the relation itself. 

It should be noted that this mass estimate, opportunely re-scaled to  $M_{500,\mathrm{opt}} = 3.5^{+1.7}_{-1.5}\times 10^{14}M_\odot$,
is smaller than the estimate based on the X-ray $Y_X$ (see Sect. \ref{section:xray_mass}), but only moderately discrepant at $2.1\sigma$ due to the large uncertainty on the optical mass estimate. We note however that \amas\ is a complex merging system for which mass estimated derived from scaling relations and calibrated mostly on relaxed systems can be largely biased. Indeed, in a non-virialized system, where the merger motions occur mainly in the plane of the sky (as it is the case for \amas), the distribution of the los-velocities cannot trace the presence of the two subclusters and the measured observed velocity dispersion might be a severe underestimate of true velocity dispersion, which is really related  to the dynamical mass. Conversely, in some phases of the merger the ICM temperature can be boosted, increasing the $Y_X$ signal, and therefore overestimating the total mass.

\begin{table}
\caption{ {\footnotesize Global optical properties of \amas.}}\label{tab:OptProp}
\begin{center}
\begin{tabular}{lr}
\toprule
Quantity            & Value     \\
\midrule
N$_{\mathrm{member\ galaxies}}$& $73$\\
RA-DEC centre$^{\mathrm{a}}$ & $179.49258; -10.76778$ [J2000] \\
RA-DEC BCG 1 &                $179.48892; -10.76686$ [J2000] \\ 
RA-DEC BCG 2 &                $179.49804; -10.76956$ [J2000] \\
$\mathrm{<z>}$  & $0.5566\pm0.0003$ \\
$\sigma_{\mathrm{V}}$ & $833_{-89}^{+103}\,\mathrm{km\ s^{-1}}$\\
R$_{200,\mathrm{opt}}$ & $1.3\pm0.1\,{\mathrm{Mpc}}$\\
M$_{200,\mathrm{opt}}$ & $4.7_{-2.0}^{+2.2}\, 10^{14}\,M_{\odot}$\\
\bottomrule
\end{tabular}
\end{center}
\footnotesize{Notes: $^{(a)}$ Cluster centre computed in \ref{sec:mem}.}
\end{table}

\subsection{Galaxy distribution and two-dimensional substructure}\label{sec:2D}

\begin{figure}[!ht]
\centering 
\includegraphics[width=9cm,trim=0 150 0 150]{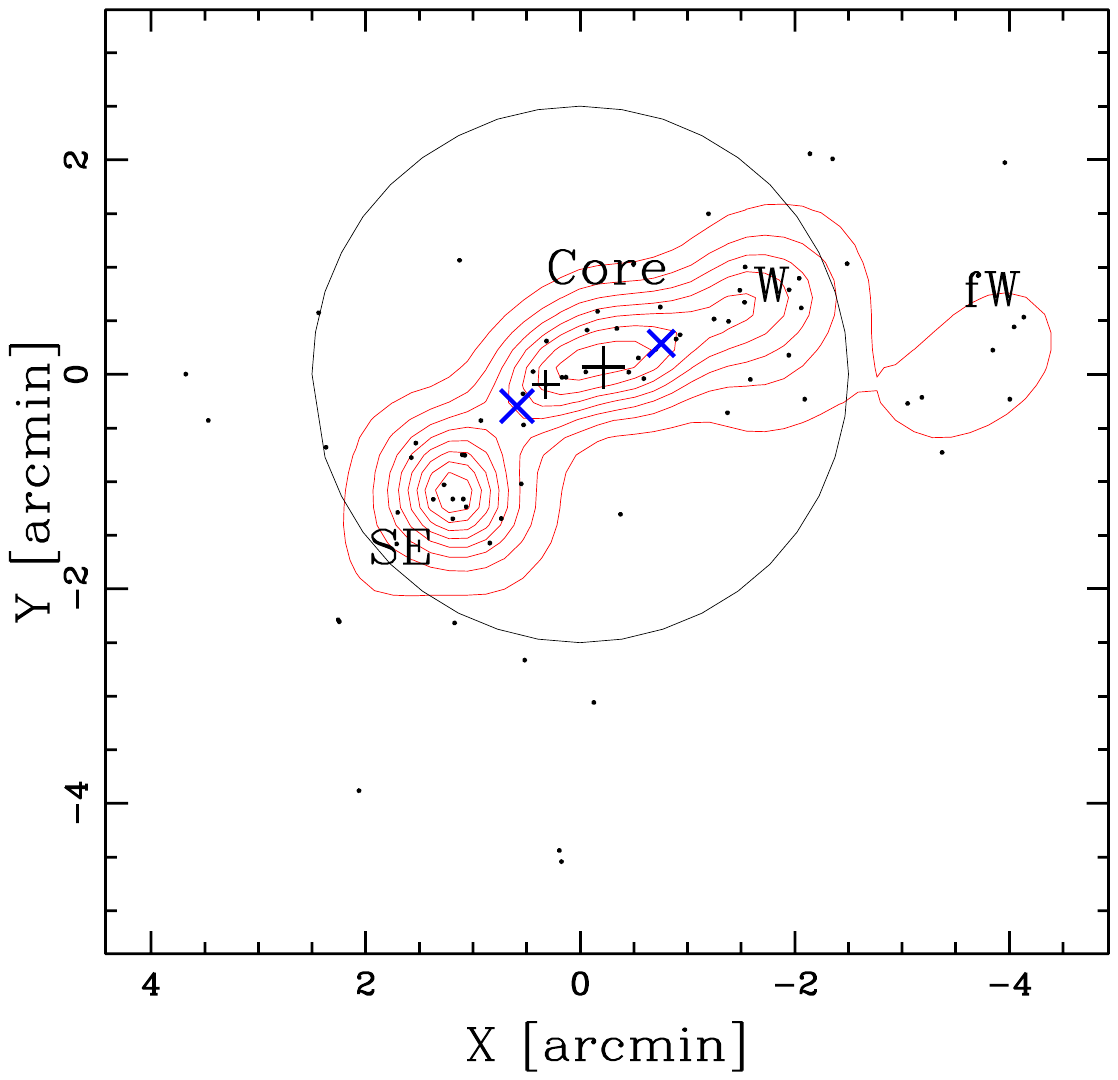}
\includegraphics[width=9cm,trim=0 150 0 150]{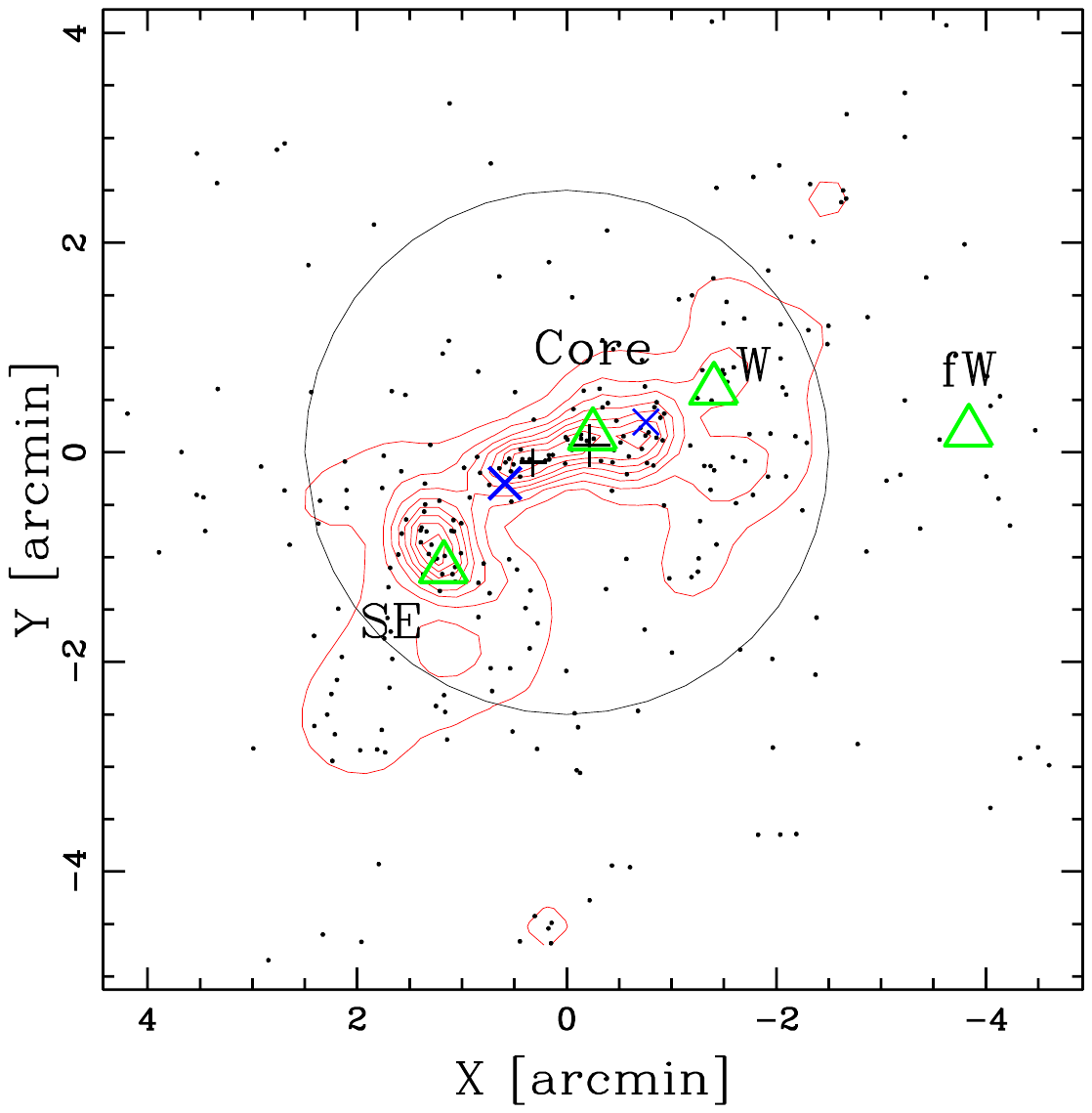}

\caption{ Spatial distribution of the galaxy members of \amas. Upper panel: Black points indicate the 73
spectroscopically selected cluster members. The red contours
represent the isodensity levels obtained with the 2D-DEDICA method
(see Table~\ref{tabdedica2dz}). The two black crosses indicate the positions of BCG1 and
BCG2. The image is centred on the cluster centre determined in
Sect. 3.2. The circle indicates a radius of 2.5 arcmin corresponding to $\sim 1$
Mpc. The blue ‘X’ symbols indicate the two cluster X-ray peaks. Lower
panel: As above, but for the 281 likely TNG photometric members (see
text).  To allow for an easy comparison, green triangles indicate the positions of the peaks in the spectroscopic sample shown in the upper panels.
}
\label{fig:2D_optical}
\end{figure}

We analysed the spatial distribution of the 73 spectroscopic member galaxies using the 2D adaptive-kernel method, hereafter 2D-DEDICA (\citealt{pisani1996}; see also \citealt{girardi1996}). The identified member galaxies are highlighted in Fig.~\ref{fig:optical_image} and  Fig.~\ref{fig:2D_optical}. 
Overall, the spatial distribution of the member galaxies is characterised by an elongated structure exhibiting substructures. In particular, we identified four clumps of galaxies detected with a c.l. higher than 99\% and distributed over the elongation direction from SE-NW. These are indicated in Fig.~\ref{fig:2D_optical} and are located in the cluster core (Core) and in the SE, west (W), and far west (fW). For each of these four clumps of galaxies, Table~\ref{tabdedica2dz} contains the number of member galaxies, the position of the 2D density peak, and its relative density, $\rho$, with respect to the densest peak. Both BCG1 and BCG2 belong to the Core. The SE peak is very populated but lacks a prominent galaxy, since its brightest galaxy has a magnitude of r$>$20 and lies at the border of the clump \ccor{($\sim 0.6$ Mpc from the SE centre, see galaxy ID.~84 in Fig.~\ref{fig:optical_image})}. The SE and Core clumps are comparable in terms of the number of members and densities. Moreover, the respective velocity dispersions are 600 \ks and 790 \kss, comparable within the large uncertainties. The SE mean velocity is consistent with the one of the Core, suggesting a merger occurring mainly in the plane of the sky.  
Conversely, the western clumps show a significant velocity deviation with respect to the rest of the cluster and they are less relevant  substructures. In fact,
these substructures are characterised by a relative density smaller
than the Core and SE-peak in the redshift sample (see
Table~\ref{tabdedica2dz}). Moreover, in the photometric samples (see below), the W-peak is detected
with an even lower relative density ($\rho\sim 0.35$) and the fW-peak
is detected with a very small relative density ($\rho<0.1$) or not
detected.

The spectroscopic sample is affected by magnitude incompleteness due to constraints on the design of the TNG and VLT MOS masks. Our TNG photometric data can help us to overcome our incompleteness problems. We selected likely members on the basis of the $r$--$i$ versus $r$ colour-magnitude relation (CMR), which indicates the locus of cluster galaxies (e.g. \citealt{goto2002}).  To determine the CMR, we applied the 2$\sigma$-clipping fitting procedure to the 73 cluster members (e.g. \citealt{boschin2012}). We obtained $r$--$i$ = $(1.302\pm0.133)-(0.008\pm0.006)\times r$, based on 53
surviving galaxies.  Out of the TNG photometric catalog, we consider as likely cluster members the objects lying within 0.15 mag from the CMR; that is, about one time the error on the intercept.
We analysed the distribution of the 281 photometric
likely members that have $r \le 24$; that is, that are $\sim 3$ mags fainter than
$M^*_r$, the characteristic absolute magnitude of the luminosity
function of galaxies in clusters.  We confirm the results obtained in
the spectroscopic sample. As above, we detect a very dense clump in the SE ($\rho=1$) with a nominal difference in position
of about 13 arcsec, corresponding to $~0.08$ Mpc at the cluster
redshift.  As for the clumps in the western region, the W-peak is
still detected with a c.l. higher than 99\% but with a lower relative
density, $\rho=0.33$. The fW-peak is only 98\% significant, with a very
small density, $\rho<0.1$.  As in the spectroscopic sample, the Core
structure appears elongated in the ESE-WNW direction. It is also
split into three peaks but without a spectroscopic confirmation we
think that it is not appropriate to emphasise this point.
\begin{table}[!ht]
        \caption{Substructure according to the 2D-DEDICA analysis of the spectroscopic member sample.}
         \label{tabdedica2dz}
            $$
         \begin{array}{l r c c c}
            \hline
            \noalign{\smallskip}
            \hline
            \noalign{\smallskip}
\mathrm{Group} & N_{\mathrm{gal}} & \alpha,\delta\,({\mathrm{J}}2000)&\rho&<V>\\
& & \mathrm{h:m:s,\degree:\arcmm:\arcs}&&\mathrm{km\ s^{-1}}\\
         \hline
         \noalign{\smallskip}
\mathrm{SE}     & 24&11\ 58\ 03.0,-10\ 47\ 11&1.00&167171\pm126 \\ 
\mathrm{Core}   & 22&11\ 57\ 57.2,-10\ 45\ 55&0.84&167000\pm174 \\ 
\mathrm{W}      & 14&11\ 57\ 52.5,-10\ 45\ 29&0.59&165786\pm307 \\ 
\mathrm{fW}     &  8&11\ 57\ 42.6,-10\ 45\ 53&0.16&165736\pm278 \\ 
\hline
              \noalign{\smallskip}
              \noalign{\smallskip}
            \noalign{\smallskip}
         \end{array}
$$

\end{table}
\begin{figure}
\centering 
\includegraphics[width=9cm,trim=0 150 0 150]{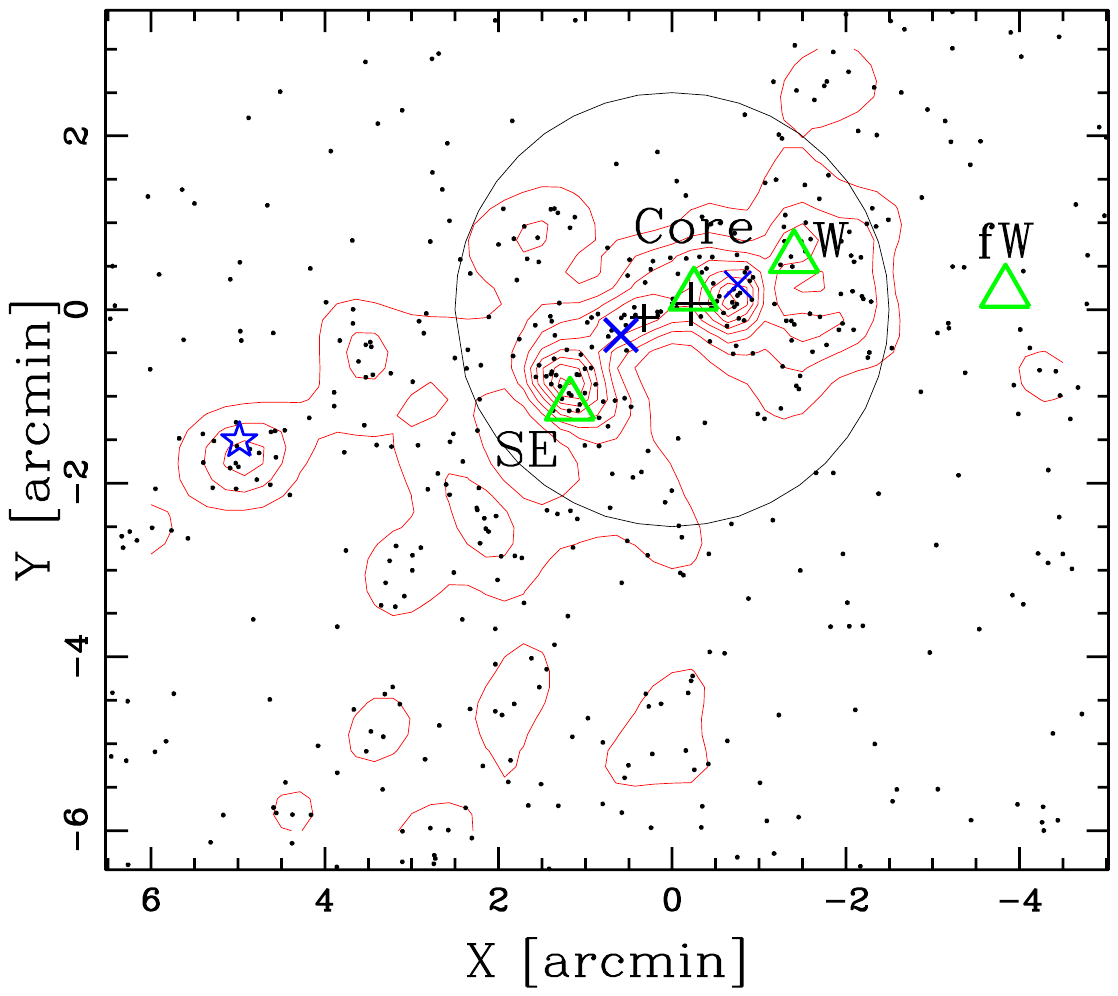}
\caption{Same as Fig.~\ref{fig:2D_optical} but for the DECaLS photometric members (see text). Green triangles indicate the positions of the peaks in the
spectroscopic sample in such a way as to allow for an easy comparison. The blue star indicates
the external ESE X-ray emission.}
\label{fig:2D_opticaldecals}
\end{figure}

In order to analyse \amas\ out to very external regions, we also considered the photometric information from the DECaLS survey and used the above technique to extract photometric members in a region of 30\arcmin x15\arcmin enclosing the cluster.    
Figure~\ref{fig:2D_opticaldecals} shows the resulting 2D-DEDICA contours, only  in the external regions for the sake of clarity. \amas\ is clearly more extended towards the east external region. The ESE extension connects the cluster with
the ESE X-ray emission (see Sect.~\ref{sec:extended_sources}) in which a galaxy concentration is present, supporting the existence of a galaxy group.
The SE filamentary extension goes farther than the SE substructure, out to 
$\lesssim 2$ Mpc from the cluster centre. There, projected to this SE cluster galaxy overdensity, is the SE X-ray emission and the likely group at $z\sim 0.19$, which we confirm using a photometric selection typical of this nearby redshift.\\
In March 2024, we were able to take new TNG spectra of the dominant galaxies of the ESE and SE galaxy overdensities and compute their spectroscopic redshifts. We found that $\mathrm{z_{spec}}=0.5521 \pm 0.0002$ and 0.1735 $\pm$ 0.0001, respectively. Our measurements definitely confirm that the ESE galaxy concentration is connected to \amas , while the SE X-ray emission is produced by the ICM of a foreground group. Nevertheless, our results seem to confirm a complex merger with at least two directions of accretion.
\subsection{Three-dimensional reconstruction}\label{sec:3D}

\begin{figure}
\centering 
\includegraphics[width=9cm,trim=0 150 0 150]{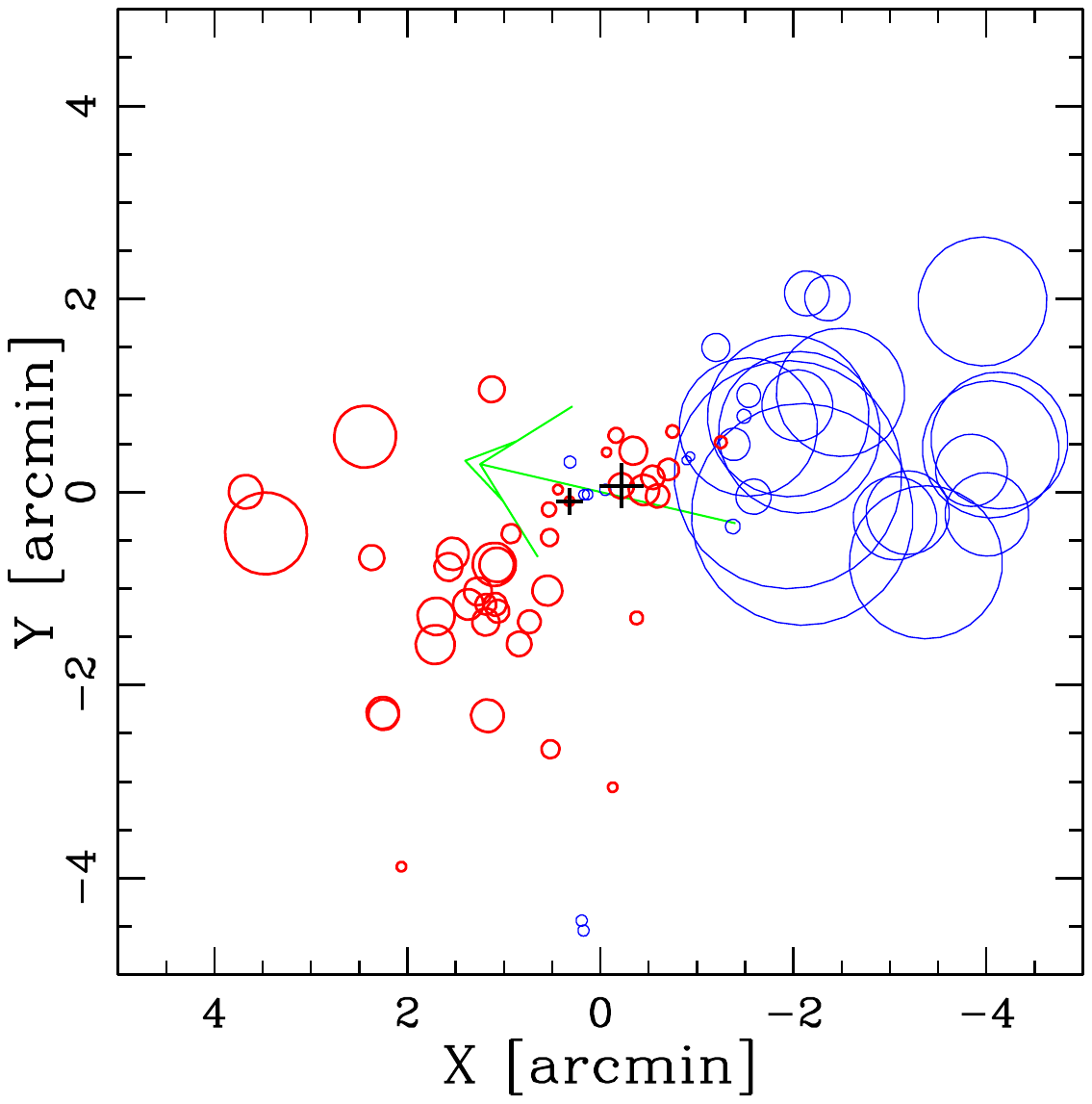}
\caption
{Dressler and Schectman bubble plot for the DSV test. Each circle indicates the position in the sky of the 73 cluster members and its radius is proportional to the deviation, $|\delta_{i,{\mathrm{V}}}|$, of the local mean velocity from the global mean velocity. The thin blue and thick red circles indicate where the local mean velocity is smaller or larger than the global mean velocity. The green arrow indicates the direction of the velocity gradient.  The diagram is centred on the centre of the cluster and the large and small crosses indicate BCG1 and BCG2, respectively.}
\label{figds10v}
\end{figure}

\ccob{As for the full 3D analysis, we looked for a correlation between the los-velocity and position information. 
The presence of an los-velocity gradient was quantified by a multiple linear regression fit to the observed velocities with respect to the galaxy positions in the plane of the sky (see also \citealt{hartog1996}). The direction of the velocity gradient on the celestial sphere is given by the angle $PA = 77_{-19}^{+23}$ degrees, measured counterclockwise from the north, which means that high-los-velocity galaxies are located in the eastern region of the cluster and low-los-velocity galaxies in the western region (see Fig.~\ref{figds10v}).}  To assess the significance of this
velocity gradient, we performed 1000 Monte Carlo simulations of clusters by randomly shuffling the velocities of the galaxies. For each simulation, we determined the coefficient of multiple determination ($RC^2$, \citealt{nag_book}).
The significance of the velocity gradient is the fraction of cases in which the $RC^2$ of the simulated data is smaller than the observed $RC^2$.  In \amas, the velocity gradient is significant at the $99.7\%$ c.l..

We used the classical $\Delta$-test of \cite{dressler1988}, from now on DS test.  For each $i$-th galaxy, the deviation of the local mean velocity from the global velocity is defined as $|\delta_{i}|$ with $\delta_{i}^2 = [(N_{\mathrm{nn}}+1)/\sigma^2_{\mathrm{V}}]\times [(\left<V\right>_{\mathrm{loc}}-\left<V\right>)^2+(\sigma_{\mathrm{V,loc}}-\sigma_{\mathrm{V}})^2]$, where the local mean velocity, $\left<V\right>_{\mathrm{loc}}$, and velocity dispersion, $\sigma_{\mathrm{V,loc}}$, were calculated using the $i$-th galaxy and its $N_{\mathrm{nn}} = 10$ neighbours. For a cluster, the cumulative deviation is given by the value of $\Delta$, which is the sum of the $|\delta_i|$ values of the individual $N$ galaxies.  We also used the modified version that considers only the indicator of local mean velocity; that is, $\delta_{i,{\mathrm{V}}} =[(N_{\mathrm{nn}}+1)^{1/2}/\sigma_{\mathrm{V}}]\times (\left< V\right>_{\mathrm{loc}} -\left<V\right>)$, and $\Delta$ is the sum of the $|\delta_{i,{\mathrm{V}}}|$ values of the individual $N$ galaxies (DSV test, e.g. \citealt{girardi2015}). As in the calculation of the velocity gradient, the significance of the $\Delta$ —  the presence of substructure — is based on 1000 Monte Carlo simulated clusters. In
\amas, the significance of the substructure is $> 99.9\%$ c.l. and $99.3\%$ c.l. according to DS and DSV tests, respectively.  In Fig.~\ref{figds10v}, we show the Dressler and Schectman bubble plot resulting from the indicator of the DSV test, $|\delta_{i,{\mathrm{V}}}|$. This plot shows very clearly how the western region is populated by
galaxies with lower velocities.

Since the DS test suggests that the velocity gradient is due to the western external structures, we repeated the analysis excluding them. We considered the system to be formed of the 46 galaxies of the Core and SE groups. No significant deviations in local velocities are detected, even when using the DS test
with five neighbours instead of ten. Accordingly, no significant velocity gradient
is detected. This suggests that any merger concerning the Core and SE clumps
is lying on the plane of the sky.

We also used the 3D-DEDICA method \citep{pisani1996,bardelli1998}. We found two galaxy peaks that are detected
with a probability higher than the $>99\%$ c.l.. The first group is
the SE group already detected by 2D-DEDICA and the second is formed by
19 galaxies, 14 of which are in common with Core group. This reinforces
our idea that the western structures are less relevant with respect to Core and SE groups.

Finally, we applied the method described by \citet{serna1996}, which is referred to in the literature as the
Serna-Gerbal or Htree method (see also e.g. \citealt{durret2009}, \citealt{girardi2011}, and \citealt{guennou2014}).  We used only the 72 member galaxies that have $r$ magnitudes.  The H-tree method uses a hierarchical cluster analysis to
determine the relationship between galaxies based on their relative
binding energy. The ignorance of the mass associated with each galaxy
is overcome by assuming the typical mass-to-light of galaxy clusters
for each galaxy halo; we assumed $M/L_r = 200$ \ml (e.g. \citealt{popesso2005}).
At low energies, the cluster splits into two groups, formed of
14 and eight galaxies, respectively.  The first group traces the core
region and, in particular, BCG1 and BCG2 lie at the bottom of the
potential well. The second group traces the SE region and
hosts the three galaxies with IDs~63, 65, and 70 at the lowest binding
energies (see also Fig.~\ref{fig:optical_image}).


\section{Merging picture}\label{sec:section4}
Our analysis of the new \chandra\ and TNG observations confirms that \amas\ is a system undergoing a major merger. However, the interpretation of our results to derive the merging scenario is not straightforward. We identify four spectroscopically confirmed clumps of galaxies indicated in Fig. \ref{fig:2D_optical}, the Core and SE being the principal galaxy concentrations. In the western external region of the cluster, the galaxy substructures W and fW  are moving along the line of sight, as is shown by our 3D analysis (Fig. \ref{figds10v}). We do not detect X-ray emission from these regions and assume that they just started their interaction with the cluster without affecting the main cluster region.\\

Moving to the central part of the cluster, the clump of cluster galaxies located SE of the X-ray peak is ahead of the gas component if we assume a motion in the SE direction. This clump does not show a clear dominant galaxy. The central main galaxy concentration traces the same ESE-WNW direction as X-ray isophotes and shows two BCGs, neither of which is associated with the primary or secondary X-ray peaks, as is shown in the top right panel of Fig. \ref{fig:wavelet_and_extended} and in Fig.~\ref{fig:optical_image}. 
The two main galaxy concentrations (Core and SE) have comparable los-velocities and no velocity gradient can be detected (see Table~\ref{tabdedica2dz} and Sect.~\ref{sec:3D}), suggesting that their merger is occurring mostly in the plane of the sky. 
The X-ray image shows a main brightness peak in the SE region associated with cooler gas (bottom right panel in Fig. \ref{fig:wavelet_and_extended} and Table \ref{tab:temperature_check}). The X-ray extension to the WNW could be interpreted as a tail of emission from the main peak, but also as a secondary cluster component, with its peak of X-ray emission. The temperature map in the bottom right panel of Fig. \ref{fig:wavelet_and_extended} shows some hints of hotter gas between the two X-ray peaks, but the variation is not significant (Table \ref{tab:temperature_check}). \\

Some properties of the central part of the cluster (Core and SE clump) are overall in agreement with the early scenario of a bimodal cluster merger, soon after core passage. In this scenario, the baryonic components of a subcluster moving in the SE direction are now separated and correspond to the SE clump of collisionless galaxies and to the X-ray peak (the collisional ICM). They feature a large separation in the plane of the sky of $\sim 350$ kpc, one of the largest observed so far \citep[see the sample in ][]{harvey2015}.
The main galaxy component at the centre could be associated with the secondary X-ray peak and the merger would be occurring close to the plane of the sky. This scenario would make \amas\ an analogue of the Bullet Cluster. However, the presence of two BCGs, neither of which is clearly associated with the secondary X-ray peak, does not easily fit this simple interpretation.
Moreover, in a Bullet Cluster-like scenario, there should be a separation between the secondary X-ray peak and the galaxies in the NW, similar to what is observed in the SE but in the opposite direction (i.e. galaxies should be more displaced towards the NW than the ICM), while in \amas\ the secondary X-ray peak is located NW of the main galaxy concentration and of BCG1.

An alternative bimodal scenario for the central part could be that we are observing two halos just before they collide approximately along the plane of the sky. The temperature map shown in the bottom right panel of Fig. \ref{fig:wavelet_and_extended} supports this scenario, in which the primary X-ray peak corresponds to the cold core of the main cluster, which is merging with the secondary cluster identified by the second X-ray peak and the galaxy concentration around BCG1. The hint of a hot region, B, may suggest that the gas is being adiabatically heated by the compression of the two in-falling objects. 
In this scenario, BCG2 should be associated with the main X-ray peak and BCG1 with the secondary peak, and they are currently located ahead of their gas components in the presumed direction of motion, as is shown in the top panels of Fig. \ref{fig:wavelet_and_extended} and in Fig.~\ref{fig:optical_image}. Support for this scenario comes from the fact that the core galaxy distribution  and the line connecting the two BCGs trace the same direction as the X-ray isophotes. On the other hand, we notice that only BCG2 is elongated in the same direction, but not BCG1.
Moreover, the role of the SE galaxy clump is not clear in this scenario: it should be a substructure of the main cluster, implying that it has already undergone a merger. \\

It is also possible that \amas\ is undergoing a more complex multiple merger than the original bimodal scenario that we envisaged from early data,
involving the SE galaxy clump, one halo associated with BCG1 and one with BCG2, moving in the plane of the sky, plus the western galaxy clumps moving along the line of sight.

The analysis of the data presented in this paper did not allow us to provide a clean merging scenario for \amas. 
A crucial role in understanding the history of the system could be played by the analysis of HST data with strong and weak gravitational lensing. This will allow one to map the overall distribution of matter and highlight the presence of DM clumps and their association with baryonic features. 
Preliminary radio analysis shows the presence of a relic to the east of the main X-ray peak as well as hints of a candidate counter relic on the opposite side of the cluster. If a double relic system is confirmed, this will give fundamental insights into the geometry of the merger.

\section{Conclusions}\label{sec:section5}
 We have studied the morphological and dynamical properties of the ICM and galaxies of \amas, which we dubbed the \planck\ bullet cluster. Our main findings on this unique object are the following:
\begin{itemize}
    \item the ICM morphology indicates that the cluster is undergoing a major merging activity. We identified the presence of a main halo and a tail, within which we determined the presence of a secondary X-ray peak. We also identified two extended emissions in the east sector that correspond to a group that is likely interacting with the cluster and a serendipitous foreground galaxy group at z$\sim$0.17;
    \item the spatially resolved spectroscopical analysis indicates the presence of a cold spot, $\sim 4$ keV, within the central parts of the main halo and a large hot region, $\sim 8-10$ keV, which corresponds spatially to the tail. The large errors on this latter region prevent us from drawing firm conclusions, but the presence of such hot regions corroborates the merging scenario;

    \item  the cluster spectroscopic redshift is $z=0.5566$. We  report the first estimate of velocity dispersion, $\sigma_{\mathrm{V}}\sim 830$ km/s;
     
     \item  \amas\ is a strongly substructured cluster showing two main galaxy concentrations $\sim 0.75$ Mpc apart at the cluster redshift
and no significant different velocity along the line of sight. 
The elongated core region and the two BCGs trace the
same WNW-ESE direction of X-ray isophotes. The second galaxy concentration
lies  towards the SE direction and does not contain a dominant galaxy;

     \item  in the western regions, we detect two minor concentrations that are
populated by low-velocity galaxies with a velocity difference of $\Delta V_{\mathrm{rf}}\sim 850$ km/s in the cluster rest frame. These structures cause the detection of a significant velocity gradient and 3D substructure;

    \item we measured the mass of \amas\ using the X-ray data via the Yx proxy and the optical data through the velocity dispersion and found $\Mvyx=7.35^{+0.64}_{-0.56} \times 10^{14}M_{\odot}$ and M$_{500,\mathrm{opt}} = 3.5^{+1.7}_{-1.5}\times 10^{14}$M$_\odot$, respectively. The difference between these masses is at $2.1\sigma$. Generally speaking, measuring the mass of such complicated objects (e.g. \citealt{furtak2024}) is challenging and potentially affected by biases; for example, the X-ray mass is calibrated on hydrostatic masses. Furthermore, the X-ray and the optical data are probing different components that are in diverse stages of virialisation. For these reasons, only WL measurements will provide firm constraints on the total mass.
 \end{itemize}
The merging picture we propose is that \amas\ is undergoing a complex major merger with at least two components moving almost in the plane of the sky, probably just after the core passage. This could have caused an offset of $\sim 350$ kpc between the galaxy and gas components, one of the largest observed so far. Despite all the results we were able to achieve, the overall picture is not complete and still presents some puzzling aspects.
The presence of multiple galaxy clumps points towards a complex scenario in which there are multiple mergers. The X-ray data are potentially seeing only the two most massive components and the mergers aligned favourably with the line of sight.
Another possible although less likely merging scenario is that we are observing two halos just before the merger and that the hot region visible in the  \chandra\ temperature map corresponding to the tail has been heated through adiabatic compression. Forthcoming radio and WL analysis will be fundamental to getting a more complete view of the complexity of this object. 

\begin{acknowledgements} 
We thank the anonymous referee for the helpful comments that improved the
quality of the manuscript. We acknowledge financial contribution from the INAF GO Programme ``Properties of the dark matter with the Planck Bullet Cluster'' and
from the European Union’s Horizon 2020 Programme under the AHEAD2020 project (grant agreement n. 871158). 
Basic research in radio astronomy at the Naval Research Laboratory is supported by 6.1 Base funding.
AZ acknowledges support by Grant No. 2020750 from the United States-Israel Binational Science Foundation (BSF) and Grant No. 2109066 from the United States National Science Foundation (NSF); by the Ministry of Science \& Technology, Israel; and by the Israel Science Foundation Grant No. 864/23.
 \end{acknowledgements}
\bibliographystyle{aa}
\bibliography{lib_articoli}

\begin{thebibliography}{69}
\expandafter\ifx\csname natexlab\endcsname\relax\def\natexlab#1{#1}\fi

\bibitem[{{Adam} {et~al.}(2017){Adam}, {Bartalucci}, {Pratt}, {Ade}, {Andr{\'e}}, {Arnaud}, {Beelen}, {Beno{\^\i}t}, {Bideaud}, {Billot}, {Bourdin}, {Bourrion}, {Calvo}, {Catalano}, {Coiffard}, {Comis}, {D'Addabbo}, {De Petris}, {D{\'e}mocl{\`e}s}, {D{\'e}sert}, {Doyle}, {Egami}, {Ferrari}, {Goupy}, {Kramer}, {Lagache}, {Leclercq}, {Mac{\'\i}as-P{\'e}rez}, {Maurogordato}, {Mauskopf}, {Mayet}, {Monfardini}, {Mroczkowski}, {Pajot}, {Pascale}, {Perotto}, {Pisano}, {Pointecouteau}, {Ponthieu}, {Rev{\'e}ret}, {Ritacco}, {Rodriguez}, {Romero}, {Ruppin}, {Schuster}, {Sievers}, {Triqueneaux}, {Tucker}, {Zemcov}, \& {Zylka}}]{adam2017}
{Adam}, R., {Bartalucci}, I., {Pratt}, G.~W., {et~al.} 2017, \aap, 598, A115

\bibitem[{{Arnaud}(1996)}]{arnaud1996}
{Arnaud}, K.~A. 1996, in Astronomical Society of the Pacific Conference Series, Vol. 101, Astronomical Data Analysis Software and Systems V, ed. G.~H. {Jacoby} \& J.~{Barnes}, 17

\bibitem[{{Arnaud} {et~al.}(2010){Arnaud}, {Pratt}, {Piffaretti}, {B{\"o}hringer}, {Croston}, \& {Pointecouteau}}]{arnaud2010}
{Arnaud}, M., {Pratt}, G.~W., {Piffaretti}, R., {et~al.} 2010, \aap, 517, A92

\bibitem[{{Bardelli} {et~al.}(1998){Bardelli}, {Pisani}, {Ramella}, {Zucca}, \& {Zamorani}}]{bardelli1998}
{Bardelli}, S., {Pisani}, A., {Ramella}, M., {Zucca}, E., \& {Zamorani}, G. 1998, \mnras, 300, 589

\bibitem[{{Bartalucci} {et~al.}(2017){Bartalucci}, {Arnaud}, {Pratt}, {D{\'e}mocl{\`e}s}, {van der Burg}, \& {Mazzotta}}]{bartalucci2017}
{Bartalucci}, I., {Arnaud}, M., {Pratt}, G.~W., {et~al.} 2017, \aap, 598, A61

\bibitem[{{Bartalucci} {et~al.}(2014){Bartalucci}, {Mazzotta}, {Bourdin}, \& {Vikhlinin}}]{bartalucci14}
{Bartalucci}, I., {Mazzotta}, P., {Bourdin}, H., \& {Vikhlinin}, A. 2014, \aap, 566, A25

\bibitem[{{Beers} {et~al.}(1990){Beers}, {Flynn}, \& {Gebhardt}}]{beers1990}
{Beers}, T.~C., {Flynn}, K., \& {Gebhardt}, K. 1990, \aj, 100, 32

\bibitem[{{Bertin} \& {Arnouts}(1996)}]{bertin1996}
{Bertin}, E. \& {Arnouts}, S. 1996, \aaps, 117, 393

\bibitem[{{Bird} \& {Beers}(1993)}]{bird1993}
{Bird}, C.~M. \& {Beers}, T.~C. 1993, \aj, 105, 1596

\bibitem[{{Bleem} {et~al.}(2020){Bleem}, {Bocquet}, {Stalder}, {Gladders}, {Ade}, {Allen}, {Anderson}, {Annis}, {Ashby}, {Austermann}, {Avila}, {Avva}, {Bayliss}, {Beall}, {Bechtol}, {Bender}, {Benson}, {Bertin}, {Bianchini}, {Blake}, {Brodwin}, {Brooks}, {Buckley-Geer}, {Burke}, {Carlstrom}, {Rosell}, {Carrasco Kind}, {Carretero}, {Chang}, {Chiang}, {Citron}, {Moran}, {Costanzi}, {Crawford}, {Crites}, {da Costa}, {de Haan}, {De Vicente}, {Desai}, {Diehl}, {Dietrich}, {Dobbs}, {Eifler}, {Everett}, {Flaugher}, {Floyd}, {Frieman}, {Gallicchio}, {Garc{\'\i}a-Bellido}, {George}, {Gerdes}, {Gilbert}, {Gruen}, {Gruendl}, {Gschwend}, {Gupta}, {Gutierrez}, {Halverson}, {Harrington}, {Henning}, {Heymans}, {Holder}, {Hollowood}, {Holzapfel}, {Honscheid}, {Hrubes}, {Huang}, {Hubmayr}, {Irwin}, {James}, {Jeltema}, {Joudaki}, {Khullar}, {Klein}, {Knox}, {Kuropatkin}, {Lee}, {Li}, {Lidman}, {Lowitz}, {MacCrann}, {Mahler}, {Maia}, {Marshall}, {McDonald}, {McMahon}, {Melchior}, {Menanteau}, {Meyer}, {Miquel}, {Mocanu},
  {Mohr}, {Montgomery}, {Nadolski}, {Natoli}, {Nibarger}, {Noble}, {Novosad}, {Padin}, {Palmese}, {Parkinson}, {Patil}, {Paz-Chinch{\'o}n}, {Plazas}, {Pryke}, {Ramachandra}, {Reichardt}, {Remolina Gonz{\'a}lez}, {Romer}, {Roodman}, {Ruhl}, {Rykoff}, {Saliwanchik}, {Sanchez}, {Saro}, {Sayre}, {Schaffer}, {Schrabback}, {Serrano}, {Sharon}, {Sievers}, {Smecher}, {Smith}, {Soares-Santos}, {Stark}, {Story}, {Suchyta}, {Tarle}, {Tucker}, {Vanderlinde}, {Veach}, {Vieira}, {Wang}, {Weller}, {Whitehorn}, {Wu}, {Yefremenko}, \& {Zhang}}]{bleem2020}
{Bleem}, L.~E., {Bocquet}, S., {Stalder}, B., {et~al.} 2020, \apjs, 247, 25

\bibitem[{{Boschin} {et~al.}(2012){Boschin}, {Girardi}, {Barrena}, \& {Nonino}}]{boschin2012}
{Boschin}, W., {Girardi}, M., {Barrena}, R., \& {Nonino}, M. 2012, \aap, 540, A43

\bibitem[{{Bourdin} \& {Mazzotta}(2008)}]{bourdin2008}
{Bourdin}, H. \& {Mazzotta}, P. 2008, \aap, 479, 307

\bibitem[{{Bourdin} {et~al.}(2013){Bourdin}, {Mazzotta}, {Markevitch}, {Giacintucci}, \& {Brunetti}}]{bourdin2013}
{Bourdin}, H., {Mazzotta}, P., {Markevitch}, M., {Giacintucci}, S., \& {Brunetti}, G. 2013, \apj, 764, 82

\bibitem[{{Brada{\v{c}}} {et~al.}(2008){Brada{\v{c}}}, {Allen}, {Treu}, {Ebeling}, {Massey}, {Morris}, {von der Linden}, \& {Applegate}}]{bradac2008}
{Brada{\v{c}}}, M., {Allen}, S.~W., {Treu}, T., {et~al.} 2008, \apj, 687, 959

\bibitem[{{Caminha} {et~al.}(2023){Caminha}, {Grillo}, {Rosati}, {Liu}, {Acebron}, {Bergamini}, {Caputi}, {Mercurio}, {Tozzi}, {Vanzella}, {Demarco}, {Frye}, {Rosani}, \& {Sharon}}]{caminha23}
{Caminha}, G.~B., {Grillo}, C., {Rosati}, P., {et~al.} 2023, \aap, 678, A3

\bibitem[{{Clowe} {et~al.}(2006){Clowe}, {Brada{\v{c}}}, {Gonzalez}, {Markevitch}, {Randall}, {Jones}, \& {Zaritsky}}]{clowe2006}
{Clowe}, D., {Brada{\v{c}}}, M., {Gonzalez}, A.~H., {et~al.} 2006, \apjl, 648, L109

\bibitem[{{Clowe} {et~al.}(2004){Clowe}, {Gonzalez}, \& {Markevitch}}]{clowe2004}
{Clowe}, D., {Gonzalez}, A., \& {Markevitch}, M. 2004, \apj, 604, 596

\bibitem[{{Croston} {et~al.}(2006){Croston}, {Arnaud}, {Pointecouteau}, \& {Pratt}}]{croston2006}
{Croston}, J.~H., {Arnaud}, M., {Pointecouteau}, E., \& {Pratt}, G.~W. 2006, \aap, 459, 1007

\bibitem[{{Danese} {et~al.}(1980){Danese}, {de Zotti}, \& {di Tullio}}]{danese1980}
{Danese}, L., {de Zotti}, G., \& {di Tullio}, G. 1980, \aap, 82, 322

\bibitem[{{Dawson} {et~al.}(2012){Dawson}, {Wittman}, {Jee}, {Gee}, {Hughes}, {Tyson}, {Schmidt}, {Thorman}, {Brada{\v{c}}}, {Miyazaki}, {Lemaux}, {Utsumi}, \& {Margoniner}}]{dawson2012}
{Dawson}, W.~A., {Wittman}, D., {Jee}, M.~J., {et~al.} 2012, \apjl, 747, L42

\bibitem[{{den Hartog} \& {Katgert}(1996)}]{hartog1996}
{den Hartog}, R. \& {Katgert}, P. 1996, \mnras, 279, 349

\bibitem[{Downers(1986)}]{nag_book}
Downers, G. 1986, NAG Fortran Workstation Handbook (IL: Numerical Algorithms Group)

\bibitem[{{Dressler} \& {Shectman}(1988)}]{dressler1988}
{Dressler}, A. \& {Shectman}, S.~A. 1988, \aj, 95, 985

\bibitem[{{Durret} {et~al.}(2009){Durret}, {Slezak}, \& {Adami}}]{durret2009}
{Durret}, F., {Slezak}, E., \& {Adami}, C. 2009, \aap, 506, 637

\bibitem[{{Eckert} {et~al.}(2017){Eckert}, {Gaspari}, {Owers}, {Roediger}, {Molendi}, {Gastaldello}, {Paltani}, {Ettori}, {Venturi}, {Rossetti}, \& {Rudnick}}]{eckert17}
{Eckert}, D., {Gaspari}, M., {Owers}, M.~S., {et~al.} 2017, \aap, 605, A25

\bibitem[{{Eckert} {et~al.}(2011){Eckert}, {Molendi}, \& {Paltani}}]{eckert2011}
{Eckert}, D., {Molendi}, S., \& {Paltani}, S. 2011, \aap, 526, A79

\bibitem[{{Fadda} {et~al.}(1996){Fadda}, {Girardi}, {Giuricin}, {Mardirossian}, \& {Mezzetti}}]{fadda1996}
{Fadda}, D., {Girardi}, M., {Giuricin}, G., {Mardirossian}, F., \& {Mezzetti}, M. 1996, \apj, 473, 670

\bibitem[{{Freeman} {et~al.}(2002){Freeman}, {Kashyap}, {Rosner}, \& {Lamb}}]{freeman2002}
{Freeman}, P.~E., {Kashyap}, V., {Rosner}, R., \& {Lamb}, D.~Q. 2002, \apjs, 138, 185

\bibitem[{{Fruscione} {et~al.}(2006){Fruscione}, {McDowell}, {Allen}, {Brickhouse}, {Burke}, {Davis}, {Durham}, {Elvis}, {Galle}, {Harris}, {Huenemoerder}, {Houck}, {Ishibashi}, {Karovska}, {Nicastro}, {Noble}, {Nowak}, {Primini}, {Siemiginowska}, {Smith}, \& {Wise}}]{fruscione2006}
{Fruscione}, A., {McDowell}, J.~C., {Allen}, G.~E., {et~al.} 2006, in \procspie, Vol. 6270, Society of Photo-Optical Instrumentation Engineers (SPIE) Conference Series, 62701V

\bibitem[{{Furtak} {et~al.}(2024){Furtak}, {Zitrin}, {Richard}, {Eckert}, {Sayers}, {Ebeling}, {Fujimoto}, {Laporte}, {Lagattuta}, {Limousin}, {Mahler}, {Meena}, {Andrade-Santos}, {Frye}, {Koekemoer}, {Kohno}, {Espada}, {Lu}, {Massey}, \& {Niemiec}}]{furtak2024}
{Furtak}, L.~J., {Zitrin}, A., {Richard}, J.~P., {et~al.} 2024, arXiv e-prints, arXiv:2404.03286

\bibitem[{{Garmire} {et~al.}(2003){Garmire}, {Bautz}, {Ford}, {Nousek}, \& {Ricker}}]{garmire2003}
{Garmire}, G.~P., {Bautz}, M.~W., {Ford}, P.~G., {Nousek}, J.~A., \& {Ricker}, Jr., G.~R. 2003, in \procspie, Vol. 4851, X-Ray and Gamma-Ray Telescopes and Instruments for Astronomy., ed. J.~E. {Truemper} \& H.~D. {Tananbaum}, 28--44

\bibitem[{{Gastaldello} {et~al.}(2014){Gastaldello}, {Limousin}, {Foex}, {Munoz}, {Verdugo}, {Motta}, {More}, {Cabanac}, {Buote}, {Eckert}, {Ettori}, {Fritz}, {Ghizzardi}, {Humphrey}, {Meneghetti}, \& {Rossetti}}]{gastaldello2014}
{Gastaldello}, F., {Limousin}, M., {Foex}, G., {et~al.} 2014, \mnras, 442, L76

\bibitem[{{Gebhardt} \& {Beers}(1991)}]{gebhardt1991}
{Gebhardt}, K. \& {Beers}, T.~C. 1991, \apj, 383, 72

\bibitem[{{Giacconi} {et~al.}(2001){Giacconi}, {Rosati}, {Tozzi}, {Nonino}, {Hasinger}, {Norman}, {Bergeron}, {Borgani}, {Gilli}, {Gilmozzi}, \& {Zheng}}]{giacconi2001}
{Giacconi}, R., {Rosati}, P., {Tozzi}, P., {et~al.} 2001, \apj, 551, 624

\bibitem[{{Girardi} {et~al.}(2011){Girardi}, {Bardelli}, {Barrena}, {Boschin}, {Gastaldello}, \& {Nonino}}]{girardi2011}
{Girardi}, M., {Bardelli}, S., {Barrena}, R., {et~al.} 2011, \aap, 536, A89

\bibitem[{{Girardi} {et~al.}(1996){Girardi}, {Fadda}, {Giuricin}, {Mardirossian}, {Mezzetti}, \& {Biviano}}]{girardi1996}
{Girardi}, M., {Fadda}, D., {Giuricin}, G., {et~al.} 1996, \apj, 457, 61

\bibitem[{{Girardi} {et~al.}(2015){Girardi}, {Mercurio}, {Balestra}, {Nonino}, {Biviano}, {Grillo}, {Rosati}, {Annunziatella}, {Demarco}, \& {Fritz}}]{girardi2015}
{Girardi}, M., {Mercurio}, A., {Balestra}, I., {et~al.} 2015, \aap, 579, A4

\bibitem[{{Goto} {et~al.}(2002){Goto}, {Sekiguchi}, {Nichol}, {Bahcall}, {Kim}, {Annis}, {Ivezi{\'c}}, {Brinkmann}, {Hennessy}, \& {Szokoly}}]{goto2002}
{Goto}, T., {Sekiguchi}, M., {Nichol}, R.~C., {et~al.} 2002, \aj, 123, 1807

\bibitem[{{Guennou} {et~al.}(2014){Guennou}, {Adami}, {Durret}, {Lima Neto}, {Ulmer}, {Clowe}, {LeBrun}, {Martinet}, {Allam}, {Annis}, {Basa}, {Benoist}, {Biviano}, {Cappi}, {Cypriano}, {Gavazzi}, {Halliday}, {Ilbert}, {Jullo}, {Just}, {Limousin}, {M{\'a}rquez}, {Mazure}, {Murphy}, {Plana}, {Rostagni}, {Russeil}, {Schirmer}, {Slezak}, {Tucker}, {Zaritsky}, \& {Ziegler}}]{guennou2014}
{Guennou}, L., {Adami}, C., {Durret}, F., {et~al.} 2014, \aap, 561, A112

\bibitem[{{Harvey} {et~al.}(2015){Harvey}, {Massey}, {Kitching}, {Taylor}, \& {Tittley}}]{harvey2015}
{Harvey}, D., {Massey}, R., {Kitching}, T., {Taylor}, A., \& {Tittley}, E. 2015, Science, 347, 1462

\bibitem[{{Hilton} {et~al.}(2021){Hilton}, {Sif{\'o}n}, {Naess}, {Madhavacheril}, {Oguri}, {Rozo}, {Rykoff}, {Abbott}, {Adhikari}, {Aguena}, {Aiola}, {Allam}, {Amodeo}, {Amon}, {Annis}, {Ansarinejad}, {Aros-Bunster}, {Austermann}, {Avila}, {Bacon}, {Battaglia}, {Beall}, {Becker}, {Bernstein}, {Bertin}, {Bhandarkar}, {Bhargava}, {Bond}, {Brooks}, {Burke}, {Calabrese}, {Carrasco Kind}, {Carretero}, {Choi}, {Choi}, {Conselice}, {da Costa}, {Costanzi}, {Crichton}, {Crowley}, {D{\"u}nner}, {Denison}, {Devlin}, {Dicker}, {Diehl}, {Dietrich}, {Doel}, {Duff}, {Duivenvoorden}, {Dunkley}, {Everett}, {Ferraro}, {Ferrero}, {Fert{\'e}}, {Flaugher}, {Frieman}, {Gallardo}, {Garc{\'\i}a-Bellido}, {Gaztanaga}, {Gerdes}, {Giles}, {Golec}, {Gralla}, {Grandis}, {Gruen}, {Gruendl}, {Gschwend}, {Gutierrez}, {Han}, {Hartley}, {Hasselfield}, {Hill}, {Hilton}, {Hincks}, {Hinton}, {Ho}, {Honscheid}, {Hoyle}, {Hubmayr}, {Huffenberger}, {Hughes}, {Jaelani}, {Jain}, {James}, {Jeltema}, {Kent}, {Knowles}, {Koopman}, {Kuehn}, {Lahav},
  {Lima}, {Lin}, {Lokken}, {Loubser}, {MacCrann}, {Maia}, {Marriage}, {Martin}, {McMahon}, {Melchior}, {Menanteau}, {Miquel}, {Miyatake}, {Moodley}, {Morgan}, {Mroczkowski}, {Nati}, {Newburgh}, {Niemack}, {Nishizawa}, {Ogando}, {Orlowski-Scherer}, {Page}, {Palmese}, {Partridge}, {Paz-Chinch{\'o}n}, {Phakathi}, {Plazas}, {Robertson}, {Romer}, {Carnero Rosell}, {Salatino}, {Sanchez}, {Schaan}, {Schillaci}, {Sehgal}, {Serrano}, {Shin}, {Simon}, {Smith}, {Soares-Santos}, {Spergel}, {Staggs}, {Storer}, {Suchyta}, {Swanson}, {Tarle}, {Thomas}, {To}, {Trac}, {Ullom}, {Vale}, {Van Lanen}, {Vavagiakis}, {De Vicente}, {Wilkinson}, {Wollack}, {Xu}, \& {Zhang}}]{hamilton2021}
{Hilton}, M., {Sif{\'o}n}, C., {Naess}, S., {et~al.} 2021, \apjs, 253, 3

\bibitem[{{Jee} {et~al.}(2014){Jee}, {Hughes}, {Menanteau}, {Sif{\'o}n}, {Mandelbaum}, {Barrientos}, {Infante}, \& {Ng}}]{jee2014}
{Jee}, M.~J., {Hughes}, J.~P., {Menanteau}, F., {et~al.} 2014, \apj, 785, 20

\bibitem[{{Kalberla} {et~al.}(2005){Kalberla}, {Burton}, {Hartmann}, {Arnal}, {Bajaja}, {Morras}, \& {P{\"o}ppel}}]{kalberla2005}
{Kalberla}, P.~M.~W., {Burton}, W.~B., {Hartmann}, D., {et~al.} 2005, \aap, 440, 775

\bibitem[{{Kravtsov} {et~al.}(2006){Kravtsov}, {Vikhlinin}, \& {Nagai}}]{krav2006}
{Kravtsov}, A.~V., {Vikhlinin}, A., \& {Nagai}, D. 2006, \apj, 650, 128

\bibitem[{{Kuntz} \& {Snowden}(2000)}]{kuntzsnowden2000}
{Kuntz}, K.~D. \& {Snowden}, S.~L. 2000, \apj, 543, 195

\bibitem[{{Kuntz} \& {Snowden}(2008)}]{kuntz2008}
{Kuntz}, K.~D. \& {Snowden}, S.~L. 2008, \aap, 478, 575

\bibitem[{{Le Brun} {et~al.}(2017){Le Brun}, {McCarthy}, {Schaye}, \& {Ponman}}]{lebrun2017}
{Le Brun}, A. M.~C., {McCarthy}, I.~G., {Schaye}, J., \& {Ponman}, T.~J. 2017, \mnras, 466, 4442

\bibitem[{{Lindegren} {et~al.}(2021){Lindegren}, {Klioner}, {Hern{\'a}ndez}, {Bombrun}, {Ramos-Lerate}, {Steidelm{\"u}ller}, {Bastian}, {Biermann}, {de Torres}, {Gerlach}, {Geyer}, {Hilger}, {Hobbs}, {Lammers}, {McMillan}, {Stephenson}, {Casta{\~n}eda}, {Davidson}, {Fabricius}, {Gracia-Abril}, {Portell}, {Rowell}, {Teyssier}, {Torra}, {Bartolom{\'e}}, {Clotet}, {Garralda}, {Gonz{\'a}lez-Vidal}, {Torra}, {Abbas}, {Altmann}, {Anglada Varela}, {Balaguer-N{\'u}{\~n}ez}, {Balog}, {Barache}, {Becciani}, {Bernet}, {Bertone}, {Bianchi}, {Bouquillon}, {Brown}, {Bucciarelli}, {Busonero}, {Butkevich}, {Buzzi}, {Cancelliere}, {Carlucci}, {Charlot}, {Cioni}, {Crosta}, {Crowley}, {del Peloso}, {del Pozo}, {Drimmel}, {Esquej}, {Fienga}, {Fraile}, {Gai}, {Garcia-Reinaldos}, {Guerra}, {Hambly}, {Hauser}, {Jan{\ss}en}, {Jordan}, {Kostrzewa-Rutkowska}, {Lattanzi}, {Liao}, {Licata}, {Lister}, {L{\"o}ffler}, {Marchant}, {Masip}, {Mignard}, {Mints}, {Molina}, {Mora}, {Morbidelli}, {Murphy}, {Pagani}, {Panuzzo}, {Pe{\~n}alosa
  Esteller}, {Poggio}, {Re Fiorentin}, {Riva}, {Sagrist{\`a} Sell{\'e}s}, {Sanchez Gimenez}, {Sarasso}, {Sciacca}, {Siddiqui}, {Smart}, {Souami}, {Spagna}, {Steele}, {Taris}, {Utrilla}, {van Reeven}, \& {Vecchiato}}]{lindegren2021}
{Lindegren}, L., {Klioner}, S.~A., {Hern{\'a}ndez}, J., {et~al.} 2021, \aap, 649, A2

\bibitem[{{Lovisari} {et~al.}(2015){Lovisari}, {Reiprich}, \& {Schellenberger}}]{lovisari2015}
{Lovisari}, L., {Reiprich}, T.~H., \& {Schellenberger}, G. 2015, \aap, 573, A118

\bibitem[{{Lumb} {et~al.}(2002){Lumb}, {Warwick}, {Page}, \& {De Luca}}]{lumb2002}
{Lumb}, D.~H., {Warwick}, R.~S., {Page}, M., \& {De Luca}, A. 2002, \aap, 389, 93

\bibitem[{{Markevitch}(2006)}]{markevitch2006}
{Markevitch}, M. 2006, in ESA Special Publication, Vol. 604, The X-ray Universe 2005, ed. A.~{Wilson}, 723

\bibitem[{{Markevitch} {et~al.}(2004){Markevitch}, {Gonzalez}, {Clowe}, {Vikhlinin}, {Forman}, {Jones}, {Murray}, \& {Tucker}}]{markevitch2004}
{Markevitch}, M., {Gonzalez}, A.~H., {Clowe}, D., {et~al.} 2004, \apj, 606, 819

\bibitem[{{Menanteau} {et~al.}(2012){Menanteau}, {Hughes}, {Sif{\'o}n}, {Hilton}, {Gonz{\'a}lez}, {Infante}, {Barrientos}, {Baker}, {Bond}, {Das}, {Devlin}, {Dunkley}, {Hajian}, {Hincks}, {Kosowsky}, {Marsden}, {Marriage}, {Moodley}, {Niemack}, {Nolta}, {Page}, {Reese}, {Sehgal}, {Sievers}, {Spergel}, {Staggs}, \& {Wollack}}]{menanteau2012}
{Menanteau}, F., {Hughes}, J.~P., {Sif{\'o}n}, C., {et~al.} 2012, \apj, 748, 7

\bibitem[{{Merten} {et~al.}(2011){Merten}, {Coe}, {Dupke}, {Massey}, {Zitrin}, {Cypriano}, {Okabe}, {Frye}, {Braglia}, {Jim{\'e}nez-Teja}, {Ben{\'\i}tez}, {Broadhurst}, {Rhodes}, {Meneghetti}, {Moustakas}, {Sodr{\'e}}, {Krick}, \& {Bregman}}]{merten2011}
{Merten}, J., {Coe}, D., {Dupke}, R., {et~al.} 2011, \mnras, 417, 333

\bibitem[{{Munari} {et~al.}(2013){Munari}, {Biviano}, {Borgani}, {Murante}, \& {Fabjan}}]{munari2013}
{Munari}, E., {Biviano}, A., {Borgani}, S., {Murante}, G., \& {Fabjan}, D. 2013, \mnras, 430, 2638

\bibitem[{{Pisani}(1993)}]{pisani1993}
{Pisani}, A. 1993, \mnras, 265, 706

\bibitem[{{Pisani}(1996)}]{pisani1996}
{Pisani}, A. 1996, \mnras, 278, 697

\bibitem[{{Planck Collaboration VIII}(2011)}]{planckesz}
{Planck Collaboration VIII}. 2011, \aap, 536, A8

\bibitem[{{Planck Collaboration XXVII}(2015)}]{planck_psz2}
{Planck Collaboration XXVII}. 2015, ArXiv e-prints [\eprint[arXiv]{1502.01598}]

\bibitem[{{Planck Collaboration XXVII}(2016)}]{PSZ2}
{Planck Collaboration XXVII}. 2016, \aap, 594, A27

\bibitem[{{Popesso} {et~al.}(2005){Popesso}, {Biviano}, {B{\"o}hringer}, {Romaniello}, \& {Voges}}]{popesso2005}
{Popesso}, P., {Biviano}, A., {B{\"o}hringer}, H., {Romaniello}, M., \& {Voges}, W. 2005, \aap, 433, 431

\bibitem[{{Randall} {et~al.}(2008){Randall}, {Markevitch}, {Clowe}, {Gonzalez}, \& {Brada{\v{c}}}}]{randall2008}
{Randall}, S.~W., {Markevitch}, M., {Clowe}, D., {Gonzalez}, A.~H., \& {Brada{\v{c}}}, M. 2008, \apj, 679, 1173

\bibitem[{{Rossetti} {et~al.}(2017){Rossetti}, {Gastaldello}, {Eckert}, {Della Torre}, {Pantiri}, {Cazzoletti}, \& {Molendi}}]{rossetti2017}
{Rossetti}, M., {Gastaldello}, F., {Eckert}, D., {et~al.} 2017, \mnras, 468, 1917

\bibitem[{{Rossetti} {et~al.}(2016){Rossetti}, {Gastaldello}, {Ferioli}, {Bersanelli}, {De Grandi}, {Eckert}, {Ghizzardi}, {Maino}, \& {Molendi}}]{rossetti2016}
{Rossetti}, M., {Gastaldello}, F., {Ferioli}, G., {et~al.} 2016, \mnras, 457, 4515

\bibitem[{{Serna} \& {Gerbal}(1996)}]{serna1996}
{Serna}, A. \& {Gerbal}, D. 1996, \aap, 309, 65

\bibitem[{{Shan} {et~al.}(2010){Shan}, {Qin}, \& {Zhao}}]{shan2010}
{Shan}, H.~Y., {Qin}, B., \& {Zhao}, H.~S. 2010, \mnras, 408, 1277

\bibitem[{{Sunyaev} \& {Zeldovich}(1972)}]{sz1972}
{Sunyaev}, R.~A. \& {Zeldovich}, Y.~B. 1972, Comments on Astrophysics and Space Physics, 4, 173

\bibitem[{{Tonry} \& {Davis}(1979)}]{tonry1979}
{Tonry}, J. \& {Davis}, M. 1979, \aj, 84, 1511

\bibitem[{{Wittman} {et~al.}(2023){Wittman}, {Stancioli}, {Finner}, {Bouhrik}, {van Weeren}, \& {Botteon}}]{wittman2023}
{Wittman}, D., {Stancioli}, R., {Finner}, K., {et~al.} 2023, arXiv e-prints, arXiv:2306.01715

\end{thebibliography}
\clearpage
\appendix

\section{Velocity catalog table}\label{appendix}
We report in Table \ref{catalogG282a} the velocity of the 91 
spectroscopically confirmed galaxy members of \amas.
\begin{table}
        \caption[]{Velocity catalog of 91 spectroscopically measured
          galaxies in the field of G282. IDs. 39 and 50 are BCG1 and
          BCG2, respectively.}
         \label{catalogG282a}
              $$ 
           \begin{array}{r c c c r r c}
            \hline
            \noalign{\smallskip}
            \hline
            \noalign{\smallskip}

\mathrm{ID} & \mathrm{m} & \mathrm{\alpha},\mathrm{\delta}\,(\mathrm{J}2000)  & r & V\,\,\,\,\,&\mathrm{\Delta}V& \mathrm{Source}\\
  & &   (11^{\mathrm{h}},-10^{\mathrm{o}})              & &\mathrm{(\,km}&\mathrm{s^{-1}\,)}&\\
            \hline
            \noalign{\smallskip}

01 &N & 57\ 40.49,45\ 28.0 &     19.97&  82754& 107 &\mathrm{T}\\    
02 &Y & 57\ 41.38,45\ 32.6 &     21.39& 165617& 122 &\mathrm{T}\\    
03 &N & 57\ 41.68,48\ 28.4 &     19.88&  63703& 134 &\mathrm{T}\\    
04 &Y & 57\ 41.75,45\ 38.1 &     22.15& 164801& 180 &\mathrm{T}\\    
05 &Y & 57\ 41.91,46\ 18.5 &     22.49& 165943& 135 &\mathrm{T}\\       
06 &Y & 57\ 42.10,44\ 06.2 &     21.92& 165710& 117 &\mathrm{T}\\       
07 &Y & 57\ 42.56,45\ 51.1 &     21.03& 165357&  75 &\mathrm{T}\\    
08 &Y & 57\ 44.48,46\ 48.3 &     22.11& 167628& 105 &\mathrm{T}\\    
09 &Y & 57\ 45.25,46\ 17.5 &     22.44& 164750& 130 &\mathrm{T}\\       
10 &Y & 57\ 45.79,46\ 20.9 &     22.98& 167246& 172 &\mathrm{T}\\    
11 &N & 57\ 46.57,47\ 01.5 &     21.43& 181571& 145 &\mathrm{T}\\    
12 &Y & 57\ 48.09,45\ 02.7 &     21.86& 166433&  67 &\mathrm{T}\\    
13 &Y & 57\ 48.64,44\ 04.1 &     22.01& 166089&  75 &\mathrm{T}\\        
14 &Y & 57\ 49.50,44\ 01.2 &     22.60& 165579& 177 &\mathrm{T}\\        
15 &Y & 57\ 49.70,46\ 18.5 &     20.22& 166611&  80 &\mathrm{T}\\    
16 &Y & 57\ 49.83,45\ 27.5 &     21.67& 165880&  82 &\mathrm{T}\\    
17 &Y & 57\ 49.91,45\ 10.9 &     22.03& 165736& 160 &\mathrm{T}\\        
18 &Y & 57\ 50.29,45\ 17.3 &     21.75& 163701&  82 &\mathrm{T}\\        
19 &Y & 57\ 50.30,45\ 53.9 &     22.64& 164881& 167 &\mathrm{T}\\        
20 &Y & 57\ 51.76,46\ 07.5 &     21.97& 163385&  87 &\mathrm{T}\\        
21 &Y & 57\ 51.96,45\ 04.6 &     21.34& 167645& 110 &\mathrm{T}\\        
22 &Y & 57\ 51.98,45\ 24.3 &     21.94& 166687& 107 &\mathrm{T}\\     
23 &Y & 57\ 52.16,45\ 17.6 &     20.89& 170065&  42 &\mathrm{V}\\    
24 &N & 57\ 52.40,46\ 57.4 &     20.41& 159256&  77 &\mathrm{T}\\    
25 &Y & 57\ 52.59,45\ 35.0 &     20.91& 163810&  72 &\mathrm{T}\\    
26 &Y & 57\ 52.63,46\ 26.1 &     22.80& 166578& 135 &\mathrm{T}\\        
27 &Y & 57\ 53.14,45\ 33.7 &     21.78& 166451& 112 &\mathrm{T}\\    
28 &Y & 57\ 53.35,44\ 34.8 &     20.81& 166372&  62 &\mathrm{V}\\    
29 &Y & 57\ 54.43,45\ 42.5 &     21.93& 166880& 132 &\mathrm{T}\\        
30 &Y & 57\ 54.58,45\ 44.9 &     21.66& 167971&  85 &\mathrm{T}\\    
31 &Y & 57\ 55.18,45\ 27.0 &     22.49& 167488&  97 &\mathrm{T}\\        
32 &N & 57\ 55.19,46\ 01.0 &     22.75& 201577& 145 &\mathrm{T}\\        
33 &Y & 57\ 55.35,45\ 50.6 &     22.22& 169321& 125 &\mathrm{T}\\    
34 &Y & 57\ 55.81,46\ 07.0 &     21.06& 169399& 100 &\mathrm{T}\\    
35 &Y & 57\ 56.01,45\ 55.5 &     21.54& 166665& 140 &\mathrm{T}\\        
36 &Y & 57\ 56.38,46\ 03.4 &     20.88& 164339&  62 &\mathrm{T}\\    
37 &Y & 57\ 56.69,47\ 22.9 &     21.64& 166574& 115 &\mathrm{T}\\        
38 &Y & 57\ 56.83,45\ 39.0 &     21.45& 165126&  72 &\mathrm{V}\\    
39 &Y & 57\ 57.34,46\ 00.7 &     19.58& 166551&  42 &\mathrm{V}\\    
40 &Y & 57\ 57.56,45\ 29.4 &     21.60& 167482&  92 &\mathrm{T}\\        
41 &Y & 57\ 57.70,49\ 08.2 &     21.33& 167357&  70 &\mathrm{V}\\    
42 &Y & 57\ 57.96,45\ 39.9 &     21.68& 166638&  97 &\mathrm{T}\\    
43 &Y & 57\ 58.01,46\ 03.3 &     21.43& 167464&  85 &\mathrm{T}\\    
44 &Y & 57\ 58.76,46\ 06.2 &     22.27& 168814& 135 &\mathrm{T}\\       
45 &N & 57\ 58.87,50\ 49.9 &     21.72& 226913& 100 &\mathrm{V}\\    

                        \noalign{\smallskip}                        
            \hline                                          
            \noalign{\smallskip}                            
            \hline                                          
         \end{array}
     $$ 
         \end{table}
\addtocounter{table}{-1}
\begin{table}
          \caption[ ]{Continued.}
     $$ 
           \begin{array}{r c c c r r c}
            \hline
            \noalign{\smallskip}
            \hline
            \noalign{\smallskip}

\mathrm{ID} & \mathrm{m} & \mathrm{\alpha},\mathrm{\delta}\,(\mathrm{J}2000)  & r & V\,\,\,\,\,&\mathrm{\Delta}V& \mathrm{Source}\\
  & &   (11^{\mathrm{h}},-10^{\mathrm{o}})              & &\mathrm{(\,km}&\mathrm{s^{-1}\,)}&\\

            \hline
            \noalign{\smallskip}

46 &Y &  57\ 58.91,46\ 06.3 &     21.56& 166987& 140 &\mathrm{T}\\    
47 &Y &  57\ 58.93,50\ 37.2 &     21.81& 166095&  75 &\mathrm{V}\\ 
48 &Y &  57\ 59.01,50\ 31.0 &     20.89& 166829&  95 &\mathrm{V}\\ 
49 &Y &  57\ 59.50,45\ 46.0 &     21.67& 166336& 110 &\mathrm{T}\\ 
50 &Y &  57\ 59.53,46\ 10.4 &     19.99& 166641&  55 &\mathrm{V}\\ 
51 &N &  57\ 59.66,46\ 08.6 &     20.86& 202284&  60 &\mathrm{T}\\ 
52 &Y &  58\ 00.01,46\ 03.1 &     22.22& 166496& 182 &\mathrm{T}\\     
53 &N &  58\ 00.02,46\ 23.7 &     20.29&  52735& 100 &\mathrm{V}\\ 
54 &N &  58\ 00.30,47\ 44.0 &     19.94&  51363& 100 &\mathrm{V}\\ 
55 &Y &  58\ 00.33,48\ 44.5 &     21.94& 166174& 115 &\mathrm{V}\\ 
56 &Y &  58\ 00.37,46\ 32.9 &     21.73& 167352&  67 &\mathrm{T}\\     
57 &Y &  58\ 00.39,46\ 15.5 &     21.89& 165589& 105 &\mathrm{T}\\ 
58 &Y &  58\ 00.46,47\ 05.9 &     22.25& 166980& 155 &\mathrm{T}\\ 
59 &Y &  58\ 01.22,47\ 25.3 &     22.49& 166714& 105 &\mathrm{T}\\ 
60 &Y &  58\ 01.65,47\ 39.0 &     21.83& 165426& 135 &\mathrm{T}\\ 
61 &N &  58\ 01.93,46\ 54.0 &     21.40& 203187&  60 &\mathrm{T}\\ 
62 &Y &  58\ 01.99,46\ 30.5 &     21.69& 168086&  67 &\mathrm{V}\\ 
63 &Y &  58\ 02.55,47\ 18.7 &     21.35& 167430&  92 &\mathrm{T}\\ 
64 &Y &  58\ 02.60,46\ 49.9 &     22.22& 166991& 127 &\mathrm{T}\\ 
65 &Y &  58\ 02.66,47\ 14.4 &     20.73& 167514&  52 &\mathrm{V}\\ 
66 &N &  58\ 02.68,48\ 12.1 &     20.76&  52194& 100 &\mathrm{V}\\ 
67 &Y &  58\ 02.70,46\ 49.6 &     21.97& 166454& 137 &\mathrm{T}\\ 
68 &Y &  58\ 02.80,45\ 00.8 &     20.98& 168302&  87 &\mathrm{V}\\ 
69 &Y &  58\ 02.99,48\ 23.7 &     20.76& 166881&  65 &\mathrm{T}\\ 
70 &Y &  58\ 03.06,47\ 14.4 &     20.73& 167168&  65 &\mathrm{T}\\ 
71 &Y &  58\ 03.06,47\ 25.4 &     21.31& 168507&  82 &\mathrm{V}\\ 
72 &Y &  58\ 03.39,47\ 06.4 &     21.42& 167232&  70 &\mathrm{V}\\ 
73 &Y &  58\ 03.80,47\ 14.5 &     21.48& 167346&  92 &\mathrm{T}\\ 
74 &Y &  58\ 04.46,46\ 43.1 &     21.93& 167589&  50 &\mathrm{V}\\ 
75 &Y &  58\ 04.63,46\ 51.2 &     21.92& 169007& 150 &\mathrm{T}\\ 
76 &N &  58\ 04.90,48\ 17.8 &     21.69& 206836&  57 &\mathrm{T}\\ 
77 &Y &  58\ 05.15,47\ 21.9 &     22.28& 166625& 110 &\mathrm{T}\\ 
78 &Y &  58\ 05.19,47\ 39.5 &     22.20& 167475& 110 &\mathrm{T}\\ 
79 &N &  58\ 05.28,48\ 56.4 &     20.75& 206966&  72 &\mathrm{V}\\ 
80 &N &  58\ 05.31,47\ 57.7 &     22.93& 230371& 102 &\mathrm{V}\\ 
81 &N &  58\ 05.32,47\ 53.5 &     22.86& 230371& 155 &\mathrm{V}\\ 
82 &Y &  58\ 06.62,49\ 57.5 &     20.83& 165987& 110 &\mathrm{T}\\ 
83 &N &  58\ 07.11,51\ 20.1 &     22.56& 274173& 100 &\mathrm{V}\\ 
84 &Y &  58\ 07.37,48\ 23.0 &     20.07& 167514&  80 &\mathrm{T}\\ 
85 &Y &  58\ 07.41,48\ 21.9 &     .....& 168734& 177 &\mathrm{T}\\ 
86 &N &  58\ 07.45,49\ 50.6 &     21.85& 211791& 100 &\mathrm{T}\\ 
87 &Y &  58\ 07.87,46\ 45.4 &     21.12& 170062& 102 &\mathrm{T}\\ 
88 &Y &  58\ 08.15,45\ 30.2 &     21.21& 167731&  90 &\mathrm{T}\\ 
89 &N &  58\ 08.33,46\ 15.7 &     19.84&  52267& 100 &\mathrm{T}\\ 
90 &Y &  58\ 12.34,46\ 30.4 &     22.19& 166112& 170 &\mathrm{T}\\ 
91 &Y &  58\ 13.19,46\ 04.5 &     21.92& 166466& 140 &\mathrm{T}\\

                        \noalign{\smallskip}                        
            \hline                                          
            \noalign{\smallskip}                            
            \hline                                          
         \end{array}
     $$ 
         \end{table}

\end{document}